# DIAGRAMMATIC METHODS IN STATISTICS AND BIASING IN THE LARGE-SCALE STRUCTURE OF THE UNIVERSE

Takahiko Matsubara*

*Department of Physics*
*The University of Tokyo*
*Bunkyo-ku, Tokyo 113, Japan*

*and*

*Department of Physics*
*Hiroshima University*
*Higashi-Hiroshima 724, Japan.*

## Abstract

We develop technique concerning statistical analyses of the large-scale structure of the universe in the presence of biasing in the structure formation. We formulate the diagrammatic method to calculate the correlation functions of the nonlocally biased field from a generally non-Gaussian density field. Our method is based on generalized Wiener-Hermite expansion of the density field. The present formalism has not only general applicability, but also practical significance, too. To show the effectiveness of this method, we revisit the problems on biasing that have been considered previously, i.e., various approximations for peak statistics and hierarchical structure of correlation functions of locally biased field. Further analyses which have not been possible so far on these problems can be performed by our formalism. The gravitational evolution of primordial fluctuation or other structure-forming processes can be described by nonlocal biasing, so can be treated by our new formalism in principle.

*E-mail address: matsu@yayoi.phys.s.u-tokyo.ac.jp

## 1 INTRODUCTION

The statistics of the large-scale structure of the universe is one of the widest window through which we can search for the origin and evolution of our universe. Various models on generating the primordial fluctuation of the universe predict different statistical properties of the universe. For instance, simpler inflationary models naturally predict the Gaussian initial fluctuation with scale-invariant primordial power spectrum (Guth & Pi 1982; Starobinskii 1982; Hawking 1982; Bardeen, Steinhardt & Turner 1983). On the other hand, topological defects (see, e.g., Kolb & Turner 1990; Vilenkin 1985) such as global monopoles, cosmic strings, domain walls or textures produce various non-Gaussian fluctuation as well as various non-simpler inflationary models (Allen, Grinstein & Wise 1987; Kofman & Linde 1985; Ortolan, Lucchin & Matarrese 1989; Salopek & Bond 1990; Salopek, Bond & Efstathiou 1989; Hodges et al. 1990; Salopek 1992), the explosion model (Ikeuchi 1981; Ostriker & Cowie 1981), and so on. Density fluctuation generated by various models determines the statistical properties of mass distribution of the present universe which are subjected by the subsequent gravitational evolution.

The mass distribution, however, can not be observed directly. We can directly observe the luminous cosmic objects, such as galaxies, clusters of galaxies, QSOs, X-ray clusters, etc. Generally, the number density of luminous cosmic objects of some kind is not proportional to the mass-density field and is biased tracers of mass. The relation between the number density field $\rho_L(\boldsymbol{r})$ of the luminous objects and the density field $\rho_b(\boldsymbol{x})$ of underlying mass has the essential effects on determining the mass distribution of the present universe. These two fields can be considered as nonlocal functionals of a primordial fluctuation field. If we can express the relation of those two fields, the relation generally should be a functional as $\rho_L(\boldsymbol{r}) = F_{\boldsymbol{r}}[\rho_b(\boldsymbol{x})]$. We assume this type of relation *exists* in this paper and explore the statistical consequences of this relation of biasing. This functional depends on the physical process to form the luminous objects and the primordial fluctuation of the universe.

As an example, in usual biased galaxy formation scenarios (Kaiser 1984; Davis et al. 1985; Bardeen et al. 1986; Coles 1989; Lumsden, Heavens & Peacock 1989), the galaxies form at high density peaks of primordial fluctuation. Adopting this hypothesis, the standard cold dark matter (CDM) model can explain the observations of two-point correlation function (e.g., Davis & Peebles 1983) on small scales. However, recent observations as Automatic Plate Measuring Facility (APM) survey (Maddox et al. 1990), Queen Mary and West field College Durham, Oxford and Toronto (QDOT) redshift survey (Efstathiou et al. 1990; Saunders, Rowan-Robinson & Lawrence 1992; Moore et al. 1992) suggest that the clustering of galaxies on large-scales $\gtrsim 10h^{-1}$Mpc is stronger than that of standard CDM model. It is pointed out that this discrepancy can be reconciled by introducing nonlocally biasing mechanisms as quasor-modulated galaxy formation (Babul & White 1991) or cooperative galaxy formation (Bower et al. 1993) rather than the simple biasing by peaks.

Though the biasing by peaks is a simple model to investigate, more complicated biasing is certainly present in reality. The real bias is determined by gravitational and hydrodynamical mechanisms (Cen & Ostriker 1992, 1993), but the right form of nonlocal bias is not known yet.

The local bias, represented by a function $\rho_L(\boldsymbol{r}) = f(\rho_b(\boldsymbol{r}))$ instead of a functional, is widely explored so far. Kaiser (1984) showed that the enhancement of two-point correlation function of clusters relative to that of galaxies (see, e.g., Bahcall & Soneira 1983; Klypin & Kopylov 1983; Hauser & Peebles 1973; Postman, Huchra & Geller 1992) can be explained by local biasing between galaxy distribution and cluster distribution. He adopted a model that the biased objects are homogeneously formed in the region where the value of underlying field $\delta_b(\boldsymbol{r})$ is larger than the threshold $\nu$ in units of *rms* $\sigma$ of underlying field, and then calculated the two-point correlation function of the biased objects in high $\nu$, and large-separation limit for random Gaussian underlying field. In this model, the biasing is local and represented by a step function as $\rho_B(\boldsymbol{r}) = \theta(\delta_b(\boldsymbol{r}) - \nu\sigma)$. The generalization of this result to $N$-point correlation functions or to the expressions with loosened restriction on $\nu$ and/or separation (Politzer & Wise 1984; Jensen & Szalay 1986) was performed. Further generalizations to the case in which the underlying field has non-Gaussianity (Fry 1986; Grinstein & Wise 1986; Matarrese, Lucchin & Bonometto 1986), and/or the case of general form of local biasing function which is not necessarily the step function (Szalay 1988; Borgani & Bonometto 1989, 1990). In these generalizations, however, the biasing is restricted to be local.

The biasing by peaks does not fall under the category of local bias. The average peak number density for Gaussian random field was analytically calculated (Doroshkevich 1970; Adler 1981; Bardeen et al. 1986). Some approximations for correlation functions of peaks are known (Bardeen et al. 1986; Otto, Politzer & Wise 1986; Cline et al. 1987). Average peak number density for generally non-Gaussian random fields for high-threshold limit is also known (Catelan, Lucchin & Matarrese 1988).

In this paper, we generalize those previous works from an unified point of view. The mathematical tools for investigating the relation between biasing and statistics in general situations are presented. As the general framework, the mean density and $N$-point correlation functions for biased fields are expressed as series expansions. The underlying field is not restricted to be Gaussian, and the biasing is not restricted to be local. The diagrammatic methods to give each terms in the series expansion are introduced. In §2, the generalized Wiener-Hermite functionals which play essential roles in our analysis are introduced. Average value of any functional of a field is expanded by Wiener-Hermite functionals. In §3, the diagrammatic method to calculate the mean density and $N$-point correlation functions of biased field is derived. More compact expressions or methods of calculation is given when the biasing is local or semi-local, which is defined below. The Fourier-space version of our methods are also explained. Our methods are applied to several theoretical problems in §4. The first application is not relevant to bias, but the relation to Edgeworth expansion is discussed. The Edgeworth expansion is recently used in analyses of astrophysical density fields. The second application is on peak theory. We show that our methods improve the techniques which is available so far for peak statistics. The next application is on hierarchical relations of correlation functions in the presence of local biasing which is investigated recently by Fry & Gaztañaga (1993; and references therein). We show our method complements their results. The last application is on the gravitational nonlinear evolution as a nonlocal biasing. We apply our methods to the calculation of three-point correlation function induced by gravity. Finally, we present the conclusions in §5. Appendices are devoted to the details of arguments and calculations.

The main purpose of this paper is to introduce a general framework to calculate the statistics of nonlocal biasing from non-Gaussian fields. In this paper, applications are restricted to the previously analyzed problems. Our methods are fairly general, and it is applied not only to the problem of biasing but also to the statistical analyses of large-scale structure of the universe.

## 2 NON-GAUSSIAN RANDOM FLUCTUATIONS

### 2.1 Correlation statistics

We introduce first a homogeneous random field $\alpha(\boldsymbol{x})$ with zero mean in three-dimensional space. For astrophysical applications, this random field corresponds to, e.g., the smoothed density fluctuations,

$$\delta_R(\boldsymbol{x}) = \int d^3y W_R(|\boldsymbol{x} - \boldsymbol{y}|)\delta(\boldsymbol{y}), \tag{2.1}$$

or normalized density fluctuation, $\delta_R(\boldsymbol{x})/\langle\delta_R^2\rangle^{1/2}$, etc., depending on specific applications. In the applications in this paper, $\alpha$ is always regarded as normalized density fluctuation. In the above notations, $\delta \equiv \rho/\bar{\rho} - 1$ is the density contrast and $W_R$ is the window function which cut the high frequency component higher than $1/R$. Two popular forms are the Gaussian window:

$$W_R^{(\mathrm{G})}(x) = \frac{1}{(2\pi R^2)^{3/2}} \exp\left(-\frac{x^2}{2R^2}\right), \tag{2.2}$$

and the top-hat window:

$$W_R^{(\mathrm{TH})}(x) = \frac{3}{4\pi R^3}\theta(R - x). \tag{2.3}$$

We review some correlation statistics for a general field $\alpha$ in this subsection.

The probability distribution for a field $\alpha$ is specified by a functional $\mathcal{P}[\alpha]$. Averaging $\langle\cdots\rangle$ is represented by a functional integration as

$$\langle\cdots\rangle = \int [d\alpha] \cdots \mathcal{P}[\alpha]. \tag{2.4}$$

The mean value of $N$-product of the field is called $N$-th moment:

$$\mu^{(N)}(\boldsymbol{x}_1, \ldots, \boldsymbol{x}_N) \equiv \langle\alpha(\boldsymbol{x}_1)\cdots\alpha(\boldsymbol{x}_N)\rangle. \tag{2.5}$$

These moments are generated by the following generating functional:

$$Z[J] \equiv \int [d\alpha]\mathcal{P}[\alpha] \exp\left[i\int d^3x J(\boldsymbol{x})\alpha(\boldsymbol{x})\right], \tag{2.6}$$

through the functional differentiation:

$$\mu^{(N)}(\boldsymbol{x}_1,\ldots,\boldsymbol{x}_N) = \left.\frac{(-i)^N \delta^N Z}{\delta J(\boldsymbol{x}_1)\cdots\delta J(\boldsymbol{x}_N)}\right|_{J=0}. \tag{2.7}$$

We usually use reduced moments rather than mere moments. The $N$-th reduced moment is defined by

$$\psi^{(N)}(\boldsymbol{x}_1,\ldots,\boldsymbol{x}_N) \equiv \langle\alpha(\boldsymbol{x}_1)\ldots\alpha(\boldsymbol{x}_N)\rangle_{\text{connected}}, \tag{2.8}$$

where $\langle\cdots\rangle_{\text{connected}}$ denotes "connected part" of the moment (Bertschinger 1992) removing the disconnected parts from the moment. The reduced moments are called semi-invariants or cumulants in probability theory and related to the connected Green's function in quantum field theory. If $\alpha$ is identified with the density contrast $\delta$, the $N$-point correlation function $\xi_N$ in the continuum limit in cosmology (e.g., Peebles 1980) is identical to the $N$-th reduced moment, neglecting shot noise effect. When the statistics of a field $\alpha$ is Gaussian, the reduced moments larger than or equal to third order are all vanish.

According to the celebrated "cumulant expansion theorem" (e.g., Ma 1985), reduced moments are generated by logarithm of moment-generating functional $Z[J]$:

$$\psi^{(N)}(\boldsymbol{x}_1,\ldots,\boldsymbol{x}_N) = \left.\frac{(-i)^N \delta^N \ln Z}{\delta J(\boldsymbol{x}_1)\cdots\delta J(\boldsymbol{x}_N)}\right|_{J=0} \tag{2.9}$$

where $\psi^{(0)}$ is set to zero formally. Thus,

$$Z[J] = \exp\left[\sum_{N=2}^{\infty}\frac{i^N}{N!}\int d^3x_1\cdots d^3x_N \psi^{(N)}(\boldsymbol{x}_1,\ldots,\boldsymbol{x}_N)J(\boldsymbol{x}_1)\cdots J(\boldsymbol{x}_N)\right] \tag{2.10}$$

provide a very useful way to tackle non-Gaussian random fields because it relates the probability distribution functional $\mathcal{P}[\alpha]$ and the reduced moments through the equation (2.6).

### 2.2 Generalized Wiener-Hermite functionals

We define generalized Wiener-Hermite functionals which play essential roles in our analyses below. For a given random field $\alpha(\boldsymbol{x})$, the generalized Wiener-Hermite functionals are defined by

$$\begin{aligned}\mathcal{H}^{(m)}(\boldsymbol{x}_1,\ldots,\boldsymbol{x}_m) &= \exp\left[\frac{1}{2}\int d^3x d^3y \alpha(\boldsymbol{x})\psi^{-1}(\boldsymbol{x},\boldsymbol{y})\alpha(\boldsymbol{y})\right] \\ &\times \frac{(-1)^m\delta^m}{\delta\alpha(\boldsymbol{x}_1)\cdots\delta\alpha(\boldsymbol{x}_m)}\exp\left[-\frac{1}{2}\int d^3x d^3y \alpha(\boldsymbol{x})\psi^{-1}(\boldsymbol{x},\boldsymbol{y})\alpha(\boldsymbol{y})\right],\end{aligned} \tag{2.11}$$

where $\psi^{-1}$ denotes an infinite dimensional inverse matrix of $\psi$ defined formally by

$$\int d^3y \psi^{-1}(\boldsymbol{x},\boldsymbol{y})\psi(\boldsymbol{y},\boldsymbol{z}) = \delta^3(\boldsymbol{x}-\boldsymbol{z}), \tag{2.12}$$

and $\psi$ is an arbitrary function of two points in space in general situations. In the following, we designate $\psi$ in the above definition of generalized Wiener-Hermite functionals as 2nd reduced moment $\psi^{(2)}$ defined by equation (2.8) with $N = 2$. The *generalized* Wiener-Hermite functionals are reduced to the Wiener-Hermite functionals in the special case $\psi(\boldsymbol{x}, \boldsymbol{y}) = \delta^3(\boldsymbol{x} - \boldsymbol{y})$. The following functionals defined by infinite dimensional linear combinations of equation (2.11) as

$$\mathcal{H}_{(m)}(\boldsymbol{x}_1, \ldots, \boldsymbol{x}_m) = \int d^3y_1 \cdots d^3y_m \psi(\boldsymbol{x}_1, \boldsymbol{y}_1) \cdots \psi(\boldsymbol{x}_m, \boldsymbol{y}_m) \mathcal{H}^{(m)}(\boldsymbol{y}_1, \ldots, \boldsymbol{y}_m), \qquad (2.13)$$

are convenient for our purposes below. Note that we distinguish two classes of functionals (2.11) and (2.13) by the place of $(m)$. The latter functionals are also called generalized Wiener-Hermite functionals. The generalized Wiener-Hermite functionals defined above are the natural generalization of the generalized Wiener-Hermite polynomials (e.g., Appel & de Fériet 1926) which have finite degrees of freedom rather than three dimensional continuum degree of freedom. The first several forms of $\mathcal{H}_{(m)}$ are

$$\mathcal{H}_{(0)} \equiv 1, \qquad (2.14)$$

$$\mathcal{H}_{(1)}(\boldsymbol{x}) = \alpha(\boldsymbol{x}), \qquad (2.15)$$

$$\mathcal{H}_{(2)}(\boldsymbol{x}_1, \boldsymbol{x}_2) = \alpha(\boldsymbol{x}_1)\alpha(\boldsymbol{x}_2) - \psi(\boldsymbol{x}_1, \boldsymbol{x}_2), \qquad (2.16)$$

$$\begin{aligned} \mathcal{H}_{(3)}(\boldsymbol{x}_1, \boldsymbol{x}_2, \boldsymbol{x}_3) &= \alpha(\boldsymbol{x}_1)\alpha(\boldsymbol{x}_2)\alpha(\boldsymbol{x}_3) \\ &\quad - \left[\psi(\boldsymbol{x}_1, \boldsymbol{x}_2)\alpha(\boldsymbol{x}_3) + \psi(\boldsymbol{x}_1, \boldsymbol{x}_3)\alpha(\boldsymbol{x}_2) + \psi(\boldsymbol{x}_2, \boldsymbol{x}_3)\alpha(\boldsymbol{x}_1)\right], \end{aligned} \qquad (2.17)$$

$$\begin{aligned} \mathcal{H}_{(4)}(\boldsymbol{x}_1, \boldsymbol{x}_2, \boldsymbol{x}_3, \boldsymbol{x}_4) &= \alpha(\boldsymbol{x}_1)\alpha(\boldsymbol{x}_2)\alpha(\boldsymbol{x}_3)\alpha(\boldsymbol{x}_4) \\ &\quad - [\psi(\boldsymbol{x}_1, \boldsymbol{x}_2)\alpha(\boldsymbol{x}_3)\alpha(\boldsymbol{x}_4) + \psi(\boldsymbol{x}_1, \boldsymbol{x}_3)\alpha(\boldsymbol{x}_2)\alpha(\boldsymbol{x}_4) + \psi(\boldsymbol{x}_1, \boldsymbol{x}_4)\alpha(\boldsymbol{x}_2)\alpha(\boldsymbol{x}_3) \\ &\quad + \psi(\boldsymbol{x}_2, \boldsymbol{x}_3)\alpha(\boldsymbol{x}_1)\alpha(\boldsymbol{x}_4) + \psi(\boldsymbol{x}_2, \boldsymbol{x}_4)\alpha(\boldsymbol{x}_1)\alpha(\boldsymbol{x}_3) + \psi(\boldsymbol{x}_3, \boldsymbol{x}_4)\alpha(\boldsymbol{x}_1)\alpha(\boldsymbol{x}_2)] \\ &\quad + \left[\psi(\boldsymbol{x}_1, \boldsymbol{x}_2)\psi(\boldsymbol{x}_3, \boldsymbol{x}_4) + \psi(\boldsymbol{x}_1, \boldsymbol{x}_3)\psi(\boldsymbol{x}_2, \boldsymbol{x}_4) + \psi(\boldsymbol{x}_1, \boldsymbol{x}_4)\psi(\boldsymbol{x}_2, \boldsymbol{x}_3)\right]. \end{aligned} \qquad (2.18)$$

The generalized Wiener-Hermite functionals $\mathcal{H}^{(m)}, \mathcal{H}_{(m)}$ are symmetric about its argument $\boldsymbol{x}_1, \ldots, \boldsymbol{x}_m$ and are regarded as functionals of a field $\alpha(\boldsymbol{x})$. Under linear transformation of a field $\alpha$:

$$\alpha'(\boldsymbol{x}) = \int d^3y M(\boldsymbol{x}, \boldsymbol{y})\alpha(\boldsymbol{y}), \qquad (2.19)$$

the generalized Wiener-Hermite functionals transform as follows:

$$\begin{aligned} &\mathcal{H}'^{(m)}(\boldsymbol{x}_1, \ldots, \boldsymbol{x}_m) \\ &\quad = \int d^3y_1 \cdots d^3y_m M^{-1}(\boldsymbol{x}_1, \boldsymbol{y}_1) \cdots M^{-1}(\boldsymbol{x}_m, \boldsymbol{y}_m) \mathcal{H}^{(m)}(\boldsymbol{y}_1, \ldots, \boldsymbol{y}_m), \end{aligned} \qquad (2.20)$$

$$\begin{aligned} &\mathcal{H}'_{(m)}(\boldsymbol{x}_1, \ldots, \boldsymbol{x}_m) \\ &\quad = \int d^3y_1 \cdots d^3y_m M(\boldsymbol{x}_1, \boldsymbol{y}_1) \cdots M(\boldsymbol{x}_m, \boldsymbol{y}_m) \mathcal{H}_{(m)}(\boldsymbol{y}_1, \ldots, \boldsymbol{y}_m), \end{aligned} \qquad (2.21)$$

where $M^{-1}$ is an infinite dimensional inverse matrix of $M$:

$$\int d^3y M^{-1}(\boldsymbol{x}, \boldsymbol{y}) M(\boldsymbol{y}, \boldsymbol{z}) = \delta^3(\boldsymbol{x} - \boldsymbol{z}). \qquad (2.22)$$

Thus, we can regard $\mathcal{H}^{(m)}$ as infinite dimensional covariant tensors and $\mathcal{H}_{(m)}$ as infinite dimensional contravariant tensors. In physical applications, the linear transformation (2.19) is considered as smoothing of a field $\alpha$ if $M$ is a low-pass filter or is considered as the Fourier transformation if $M(\boldsymbol{x}, \boldsymbol{y}) = \exp(-i\boldsymbol{x}\cdot\boldsymbol{y})$. The following recursion relation,

$$\left[\alpha(\boldsymbol{x}_m) - \int d^3y \psi(\boldsymbol{x}_{m+1}, \boldsymbol{y}) \frac{\delta}{\delta\alpha(\boldsymbol{y})}\right] \mathcal{H}_{(m)}(\boldsymbol{x}_1, \ldots, \boldsymbol{x}_m) = \mathcal{H}_{(m+1)}(\boldsymbol{x}_1, \ldots, \boldsymbol{x}_{m+1}), \quad (2.23)$$

are helpful properties of the functionals.

### 2.3 Expectation values

The averaging (2.4) for a non-Gaussian random field is reduced to the averaging for a Gaussian random field using the generalized Wiener-Hermite functionals as we shall see below. First of all, with a given distribution $\mathcal{P}[\alpha]$, we associate a Gaussian distribution $\mathcal{P}_{\rm G}[\alpha]$ constructed by the 2nd reduced moment $\psi$ of the original distribution $\mathcal{P}[\alpha]$:

$$P_{\rm G}[\alpha] = N_{\rm G} \exp\left[-\frac{1}{2}\int d^3x d^3y \alpha(\boldsymbol{x})\psi^{-1}(\boldsymbol{x}, \boldsymbol{y})\alpha(\boldsymbol{y})\right], \quad (2.24)$$

where $N_{\rm G}$ is a formal normalization constant:

$$N_{\rm G} = \left\{\int [d\alpha] \exp\left[-\frac{1}{2}\int d^3x d^3y \alpha(\boldsymbol{x})\psi^{-1}(\boldsymbol{x}, \boldsymbol{y})\alpha(\boldsymbol{y})\right]\right\}^{-1}. \quad (2.25)$$

The averaging by this Gaussian distribution $\mathcal{P}_{\rm G}[\alpha]$ is denoted as $\langle\cdots\rangle_{\rm G}$ distinguished from the averaging by the original non-Gaussian field: $\langle\cdots\rangle$.

The relation between the non-Gaussian distribution $\mathcal{P}[\alpha]$ and the Gaussian distribution $\mathcal{P}[\alpha]$ is

$$\mathcal{P}[\alpha] = \exp\left[\sum_{N=3}^{\infty} \frac{(-1)^N}{N!} \int d^3x_1 \cdots d^3x_N \psi^{(N)}(\boldsymbol{x}_1, \ldots, \boldsymbol{x}_N) \frac{\delta}{\delta\alpha(\boldsymbol{x}_1)} \cdots \frac{\delta}{\delta\alpha(\boldsymbol{x}_N)}\right] P_{\rm G}[\alpha]. \quad (2.26)$$

This relation is proven as follows: the generating functional $Z[J]$ is (infinite dimensional) Fourier transform of distribution functional $\mathcal{P}[\alpha]$. Thus the inverse Fourier transformation gives the expression of $\mathcal{P}[\alpha]$ in terms of reduced moments because of equation (2.10). Substituting $i\delta/\delta\alpha(\boldsymbol{x})$ for $J(\boldsymbol{x})$, and performing Gaussian integration, we finally arrive at the relation (2.26). Expanding the exponential of equation (2.26) and using the definition of the generalized Wiener-Hermite functionals, we can express non-Gaussian averaging of arbitrary functional $F[\alpha]$ by Gaussian averaging as follows:

$$\begin{aligned}\langle F[\alpha]\rangle &= \langle F[\alpha]\rangle_{\rm G} \\ &+ \sum_{m=1}^{\infty} \frac{1}{m!} \sum_{n_1=3}^{\infty} \cdots \sum_{n_m=3}^{\infty} \frac{1}{n_1! \cdots n_m!} \int \prod_{l=1}^{m} \prod_{k=1}^{n_l} d^3x_l^{(k)} \cdot \prod_{l=1}^{m} \psi^{(n_l)}(\boldsymbol{x}_l^{(1)}, \ldots \boldsymbol{x}_l^{(n_l)}) \\ &\times \left\langle \mathcal{H}^{(n_1+\cdots+n_m)}(\boldsymbol{x}_1^{(1)}, \ldots, \boldsymbol{x}_1^{(n_1)}, \ldots, \boldsymbol{x}_m^{(1)}, \ldots, \boldsymbol{x}_m^{(n_m)}) F[\alpha]\right\rangle_{\rm G}. \end{aligned} \quad (2.27)$$

This equation is an expansion of a non-Gaussian averaging by higher order reduced moments. The nontrivial part in the equation (2.27) is the factor of the form, $\langle \mathcal{H} F\rangle_{\rm G}$. In Appendix A, the methods to calculate the Gaussian averaging of the product of generalized Wiener-Hermite functionals, $\langle \prod \mathcal{H}\rangle_{\rm G}$ are developed. Therefore, if a functional $F[\alpha]$ is expressed by the sum of the product of generalized Wiener-Hermite functionals, the right-hand-side of equation (2.27) can be calculated as performed in the next section.

# 3 DIAGRAMMATIC CALCULUS OF $N$-POINT CORRELATION FUNCTIONS

## 3.1 Generalized Wiener-Hermite expansion and diagrammatic methods

In this paper, we are interested in the correlations of a density field $\rho_{\rm B}(\boldsymbol{r},[\alpha])$ which appears by some biasing mechanism from a field $\alpha(\boldsymbol{x})$. We formally do not restrict the form of biasing in this subsection and consider a given, generally nonlocal, biasing functional $\rho_{\rm B}(\boldsymbol{r},[\alpha])$ which is both a function of a point $\boldsymbol{r}$ and a functional of a field $\alpha(\boldsymbol{x})$. It is natural to expand this functional by the generalized Wiener-Hermite functionals as follows:

$$\rho_{\rm B}(\boldsymbol{r},[\alpha]) = \sum_{m=0}^{\infty} \frac{1}{m!} \int d^3x^{(1)} \cdots d^3x^{(m)} K^{(m)}(\boldsymbol{r}-\boldsymbol{x}^{(1)},\cdots,\boldsymbol{r}-\boldsymbol{x}^{(m)}) \mathcal{H}_{(m)}(\boldsymbol{x}^{(1)},\ldots,\boldsymbol{x}^{(m)}), \tag{3.1}$$

where the kernel $K^{(m)}$ is given by

$$K^{(m)}(\boldsymbol{r}-\boldsymbol{x}^{(1)},\ldots,\boldsymbol{r}-\boldsymbol{x}^{(m)}) = \left\langle \mathcal{H}^{(m)}(\boldsymbol{x}^{(1)},\ldots,\boldsymbol{x}^{(m)}) \rho_{\rm B}(\boldsymbol{r},[\alpha]) \right\rangle_{\rm G} \tag{3.2}$$

$$= \left\langle \frac{\delta^m \rho_{\rm B}(\boldsymbol{r},[\alpha])}{\delta\alpha(\boldsymbol{x}^{(1)})\cdots\delta\alpha(\boldsymbol{x}^{(m)})} \right\rangle_{\rm G}. \tag{3.3}$$

The equation (3.2) is derived from the orthogonality of the generalized Wiener-Hermite functionals (A.20). We assume that the expansion (3.1) exists for the nonlocal biasing of our interest. Using equation (2.27), the $N$-th moment of a biased density field $\rho_{\rm B}$ is given by somewhat complicated expression:

$$P_N(\boldsymbol{r}_1,\ldots,\boldsymbol{r}_N) \equiv \langle \rho_{\rm B}(\boldsymbol{r}_1,[\alpha]) \cdots \rho_{\rm B}(\boldsymbol{r}_N,[\alpha]) \rangle \tag{3.4}$$

$$= \sum_{m_1=0}^{\infty} \cdots \sum_{m_N=0}^{\infty} \int \prod_{i=1}^{N} \prod_{a=1}^{m_i} d^3x_i^{(a)} \cdot \prod_{i=1}^{N} \frac{1}{m_i!} K^{(m_i)}(\boldsymbol{r}_i - \boldsymbol{x}_i^{(1)},\ldots,\boldsymbol{r}_i - \boldsymbol{x}_i^{(m_i)})$$

$$\times \left[ \left\langle \prod_{i=1}^{N} \mathcal{H}_{m_i}(\boldsymbol{x}_i^{(1)},\ldots,\boldsymbol{x}_i^{(m_i)}) \right\rangle_{\rm G} + \sum_{m=1}^{\infty} \frac{1}{m!} \sum_{n_1=3}^{\infty} \cdots \sum_{n_m=3}^{\infty} \frac{1}{n_1!\cdots n_m!} \right.$$

$$\times \int \prod_{l=1}^{m} \prod_{k=1}^{n_l} d^3y_l^{(k)} \cdot \prod_{l=1}^{m} \psi^{(n_l)}(\boldsymbol{y}_l^{(1)},\ldots,\boldsymbol{y}_l^{(n_l)})$$

$$\times \left\langle \mathcal{H}^{(n_1+\cdots+n_m)}(\boldsymbol{y}_1^{(1)},\ldots,\boldsymbol{y}_1^{(n_1)},\ldots,\boldsymbol{y}_m^{(1)},\ldots,\boldsymbol{y}_m^{(n_m)}) \prod_{i=1}^{N} \mathcal{H}_{(m_i)}(\boldsymbol{x}_i^{(1)},\ldots,\boldsymbol{x}_i^{(m_i)}) \right\rangle_{\mathrm{G}} \Bigg]. \quad (3.5)$$

This expression can be calculated making use of the methods described in Appendix A. The diagrammatic method in calculating Gaussian expectation value of the product of generalized Wiener-Hermite functionals is translated to the diagrammatic method in calculating moments (3.5) of $\rho_{\mathrm{B}}$. The result is summarized as the following procedures:

A-i) Draw $N$ open circles corresponding to $\boldsymbol{r}_1,\ldots,\boldsymbol{r}_N$ which we call external points.

A-ii) Consider one of possible graphs with lines between external points or between external points and vertices. Vertices are the parts where three or more ends of lines are gathered. Do not connect two vertices directly with lines. Do not draw any line whose two ends are placed at the same external point or the same vertex.

A-iii) Associate labels like $\boldsymbol{x}, \boldsymbol{y}$, etc. with ends of lines placed at external points (not at vertices).

A-iv) Apply the correspondence of Figure 1 for the external points, lines between external points and vertices. Make the product of these factors of the graph.

A-v) Integrate with respect to all the labels and multiply the following statistical factor:

$$\frac{1}{m! \prod\limits_{\mathrm{ends}} a_{\mathrm{ends}}!}, \quad (3.6)$$

where $m$ is the number of vertices and $a_{\mathrm{ends}}$ is the number of lines whose ends are identical. This statistical factor is for the case we distinguish the vertices in the graph, i.e., we distinguish the graphs like Figure 13 described in Appendix B (see Appendix B for more clear understanding of this situation).

A-vi) Sum up values obtained from all the possible graphs.

The readers are invited to calculate several examples both by equation (3.5) and by the above diagrammatic rules. The Appendix B is devoted to the explanation of the statistical factor (3.6).

This diagrammatic rules enable us to easily calculate lower order terms about $\psi^{(m)}$ or $K^{(m)}$. Although we are not guaranteed the sufficient convergence of this diagrammatic expansion in the general situation, there are many cases in astrophysics where this expansion is useful as we shall see in §4.

The reduced $N$-point correlation function of the biased density field $\rho_{\mathrm{B}}$:

$$\xi_{\mathrm{B}}^{(N)}(\boldsymbol{r}_1,\ldots,\boldsymbol{r}_N) = \frac{1}{\bar{\rho}_{\mathrm{B}}^N} \left\langle \rho_{\mathrm{B}}(\boldsymbol{r}_1,[\alpha]) \cdots \rho_{\mathrm{B}}(\boldsymbol{r}_N,[\alpha]) \right\rangle_{\mathrm{connected}} \quad (3.7)$$

is more easily evaluated in this diagrammatic expansion because the contribution to the reduced moment in equation (3.7) comes from connected graphs in our approach. This especially simplify the calculation of higher order correlation functions of the biased field. Thus, the above rules are applicable to the calculation of reduced $N$-point correlation functions substituting the following rule for A-vi).

A-vi) Sum up values obtained from all the connected graphs and multiply $(P_1)^{-N}$.

### 3.2 Local and semi-nonlocal biases

Here we consider special cases of the above general consideration, i.e., the case biasing is local or semi-nonlocal.

The local bias is defined so that the biasing functional $\rho_{\rm B}(\boldsymbol{r}, [\alpha])$ is a mere function of the underlying field $\alpha$ at the position $\boldsymbol{r}$:

$$\rho_{\rm B} = \rho_{\rm B}\left(\alpha(\boldsymbol{r})\right). \tag{3.8}$$

In this case, the kernels [eq. (3.3)] have the forms,

$$K^{(m)}(\boldsymbol{r} - \boldsymbol{x}^{(1)}, \ldots, \boldsymbol{r} - \boldsymbol{x}^{(m)}) = \delta^3(\boldsymbol{r} - \boldsymbol{x}^{(1)}) \cdots \delta^3(\boldsymbol{r} - \boldsymbol{x}^{(m)}) R_m, \tag{3.9}$$

where

$$R_m = \left\langle \frac{\partial^m \rho_{\rm B}}{\partial \alpha^m} \right\rangle. \tag{3.10}$$

If the field $\alpha$ is normalized so that $\langle \alpha^2 \rangle = 1$, $R_m$'s are coefficients of the Hermite expansion of the local biasing function:

$$\rho_{\rm B}(\alpha) = \sum_{m=0}^{\infty} \frac{R_m}{m!} H_m(\alpha); \tag{3.11}$$

$$R_m = \frac{1}{\sqrt{2\pi}} \int_{-\infty}^{\infty} d\alpha e^{-\alpha^2/2} \rho_{\rm B}(\alpha) H_m(\alpha). \tag{3.12}$$

The definition of Hermite polynomials and its orthogonality relations are,

$$H_m(\alpha) = e^{\alpha^2/2} \left( -\frac{d}{d\alpha} \right)^m e^{-\alpha^2/2}, \tag{3.13}$$

and

$$\frac{1}{\sqrt{2\pi}} \int_{-\infty}^{\infty} dx e^{-x^2/2} H_m(x) H_n(x) = m!\, \delta_{mn}, \tag{3.14}$$

respectively. If the field $\alpha$ is not normalized and $\langle \alpha^2 \rangle = \sigma^2 \neq 1$, the corresponding expansion is

$$\rho_{\rm B}(\alpha) = \sum_{m=0}^{\infty} \frac{R_m}{m!} \sigma^m H_m(\alpha/\sigma); \tag{3.15}$$

$$R_m = \frac{1}{\sqrt{2\pi}\sigma^{m+1}} \int_{-\infty}^{\infty} d\alpha e^{-\alpha^2/(2\sigma^2)} \rho_{\rm B}(\alpha) H_m(\alpha/\sigma). \tag{3.16}$$

The special form of equation (3.9) reduce the diagrammatic rules A-i) $\sim$ A-vi) above to the following:

B-i) Same as A-i)

B-ii) Same as A-ii)

B-iii) Apply the correspondence of Figure 2 and make the product of these factors of the graph.

B-iv) Multiply the statistical factor (3.6).

B-v) For $P_N$, sum up values obtained from all the possible graphs. For reduced $N$-point correlation function, sum up values obtained from all the connected graphs and multiply $(P_1)^{-N}$.

The above rules for $P_N$ could be expressed by the following equation:

$$P_N(\boldsymbol{r}_1,\ldots,\boldsymbol{r}_N) = \sum_{m=0}^{\infty}\sum_{n_{11}=0}^{\infty}\cdots\sum_{n_{1N}=0}^{\infty}\cdots\sum_{n_{m1}=0}^{\infty}\cdots\sum_{n_{mN}=0}^{\infty}\frac{R_{\Sigma_j n_{j1}}\cdots R_{\Sigma_j n_{jN}}}{m!\prod_{j=1}^{m}\prod_{k=1}^{N} n_{jk}!}$$
$$w^{(\Sigma_k n_{1k})}(\underbrace{1,\ldots,1}_{n_{11}},\ldots,\underbrace{N,\ldots,N}_{n_{1N}})\cdots w^{(\Sigma_k n_{mk})}(\underbrace{1,\ldots,1}_{n_{m1}},\ldots,\underbrace{N,\ldots,N}_{n_{mN}}). \qquad (3.17)$$

where

$$\begin{aligned} &w^{(0)} = 0,\\ &w^{(1)}(i) = 0,\\ &w^{(2)}(i,j) = (1-\delta_{ij})\psi^{(2)}(\boldsymbol{r}_i,\boldsymbol{r}_j)\\ &w^{(n)}(i_1,\ldots,i_n) = \psi^{(n)}(\boldsymbol{r}_{i_1}\ldots,\boldsymbol{r}_{i_n}) \qquad (n\geq 3). \end{aligned} \qquad (3.18)$$

Identifying the underlying field $\alpha$ with normalized density fluctuation $\delta_R(\boldsymbol{x})/\langle\delta_R^2\rangle^{1/2}$, this formula is an alternative expression to one obtained by Borgani & Bonometto (1989, 1990) who adopted the different approach to treat the local biasing function $\rho_{\rm B}$ in evaluating the same quantity of equation (3.17). Our expression seems to be simpler than theirs.

When the statistics of an underlying field is random Gaussian, equation (3.17) reduces to

$$P_N(\boldsymbol{r}_1,\ldots,\boldsymbol{r}_N) = \sum_{m=0}^{\infty}\sum_{\{m_{kl}|m\}}\prod_{k<l}\frac{[\psi(\boldsymbol{r}_k,\boldsymbol{r}_l)]^{m_{kl}}}{m_{kl}!}\cdot\prod_{r=1}^{N} R_{m_r}, \qquad (3.19)$$

In the above equation, $1\leq k,l\leq N$,

$$m_r = \sum_{k=1}^{r-1} m_{kr} + \sum_{k=r+1}^{N} m_{rk} \qquad (3.20)$$

and $\sum_{\{m_{kl}|m\}}$ means the sum over $m_{kl}$'s ($k < l$) which are non-negative integers with a constraint, $\sum_{k<l} m_{kl} = m$. We can see, using multinomial expansion theorem, that the equation (3.19) is equivalent to the formula obtained by Szalay (1988).

When the local biasing function $\rho_{\rm B}$ is given by sharp clipping:

$$\rho_{\rm B}(\alpha) = C\theta(\alpha - \nu), \tag{3.21}$$

then the coefficients $R_m$ are given by

$$R_m^{(\rm sc)} = \begin{cases} \dfrac{C}{2}\mathrm{erfc}\left(\dfrac{\nu}{\sqrt{2}}\right) & (m = 0) \\ \dfrac{C}{\sqrt{2\pi}} H_{m-1}(\nu) e^{-\nu^2/2} & (m \geq 1) \end{cases}, \tag{3.22}$$

where

$$\mathrm{erfc}(x) = 1 - \mathrm{erf}(x) = \frac{2}{\sqrt{\pi}} \int_x^\infty e^{-x^2} dt \tag{3.23}$$

is a complementary error function. In this case, equation (3.17) reduces to the formula obtained by Matarrese et al. (1986). If we, furthermore, constrain the statistics of underlying field to be Gaussian, and local biasing function to be the sharp clipping, we can find equation (3.19) reduces to the the formula obtained by Jensen and Szalay (1986).

High-$\nu$ limit of equation (3.17) for sharp-clipping biasing function (3.21) enables us to sum up infinite series of graphs and results in

$$P_N = \left(\frac{C}{\sqrt{2\pi}\nu e^{\nu^2/2}}\right)^N \times \exp\left[\sum_{m=2}^{\infty} \nu^m \sum_{\substack{n_1,\ldots,n_N \geq 0 \\ n_1+\cdots+n_N=m}} \frac{1}{n_1! \cdots n_N!} w^{(m)}(\underbrace{1,\ldots,1}_{n_1},\ldots,\underbrace{N,\ldots,N}_{n_N})\right] \tag{3.24}$$

$$= \left(\frac{C}{\sqrt{2\pi}\nu e^{\nu^2}}\right)^N Z_N\left(J_i = \nu\right), \tag{3.25}$$

where

$$Z_N\left(J_i = \nu\right) = \left\langle \exp\left(\nu \sum_{i=1}^{N} \alpha(\boldsymbol{r}_i)\right)\right\rangle, \tag{3.26}$$

is a $N$-point moment generating function $Z_N(J_i)$ in which $\nu$ is substituted for all $J_i$. This formula is equivalent to the one obtained by Matarrese et al. (1986). The $N = 2$ case of equation (3.24) has the same form obtained by Grinstein & Wise (1986).

The local bias is not relevant to the biasing through density peaks (Davis et al. 1985; Bardeen et al. 1986; Coles 1989; Lumsden et al. 1989) because density peaks are described by not only the mere value of the underlying field but also spatial derivatives of it (see §4 for details). Thus, we next consider the case that the biasing functional $\rho_{\rm B}(\boldsymbol{r}, [\alpha])$ depends on $\alpha(\boldsymbol{r})$ and finite-order derivatives of it at $\boldsymbol{r}$, generally. We denote the spatial derivatives of a underlying field $\alpha$ as $\alpha_\mu$:

$$(\alpha_\mu(\boldsymbol{x})) = \left( \underbrace{\alpha(\boldsymbol{x})}_{1}, \underbrace{\frac{\partial \alpha}{\partial x^i}}_{3}, \underbrace{\frac{\partial^2 \alpha}{\partial x^i \partial x^j}(i \leq j)}_{6}, \ldots \right), \tag{3.27}$$

where $\mu = 0, 1, 2, \ldots$. This notation enable us to write the semi-nonlocal biasing functional as

$$\rho_{\mathrm{B}}(\boldsymbol{r}, [\alpha]) = \rho_{\mathrm{B}}\left(\alpha_\mu(\boldsymbol{r})\right). \tag{3.28}$$

In this case, kernels (3.3) have the forms,

$$K^{(m)}(\boldsymbol{r} - \boldsymbol{x}^{(1)}, \ldots, \boldsymbol{r} - \boldsymbol{x}^{(m)}) = \sum_{\mu_1} \cdots \sum_{\mu_m} \delta_{\mu_1}(\boldsymbol{r} - \boldsymbol{x}^{(1)}) \cdots \delta_{\mu_m}(\boldsymbol{r} - \boldsymbol{x}^{(m)}) R_{\mu_1 \cdots \mu_m}, \tag{3.29}$$

where $\delta_\mu$ denotes the derivatives of Dirac's delta-function in the notation of equation (3.27), and

$$R_{\mu_1 \cdots \mu_m} \equiv \left\langle \frac{\partial}{\partial \alpha_{\mu_1}} \cdots \frac{\partial}{\partial \alpha_{\mu_m}} \rho_{\mathrm{B}}(\alpha_\mu) \right\rangle_{\mathrm{G}}. \tag{3.30}$$

The special form of equation (3.29) reduce the diagrammatic rules A-i) $\sim$ A-vi) above to the following:

C-i) Same as A-i)

C-ii) Same as A-ii)

C-iii) Associate labels like $\mu, \nu$, etc. with ends of lines placed at external points.

C-iv) Apply the correspondence of Figure 3 and make the product of these factors of the graph.

C-v) Perform sums with respect to all the labels and multiply the statistical factor (3.6).

C-vi) For $P_N$, sum up values obtained from all the possible graphs. For reduced $N$-point correlation function, sum up values obtained from all the connected graphs and multiply $(P_1)^{-N}$.

In Figure 3, $\psi^{(2)}_{\mu\nu}(\boldsymbol{r}_i, \boldsymbol{r}_j)$ denotes the derivatives of $\psi^{(2)}$ and $\mu$ indicates the derivatives with respect to $\boldsymbol{r}_i$ according to equation (3.27), $\nu$ indicates the derivatives with respect to $\boldsymbol{r}_j$. Similar definition is applied to $\psi^{(n)}_{\mu\nu\cdots\lambda}(\boldsymbol{r}_i, \ldots, \boldsymbol{r}_k)$.

### 3.3 Fourier transforms

The biasing may be expressed in Fourier space rather than a coordinate space. For example, the transfer functions $T(k)$ of various cosmological models is a kind of biasing in our context through Fourier space. It may be useful for the analyses in the future to give the diagrammatic rules in Fourier space.

The Fourier transformation is a kind of the following linear transformations of a field:

$$\alpha'(\boldsymbol{k}) = \int d^3x M(\boldsymbol{k},\boldsymbol{x})\alpha(\boldsymbol{x}) \tag{3.31}$$

$$\rho'_{\rm B}(\boldsymbol{p},[\alpha']) = \int d^3r L(\boldsymbol{v},\boldsymbol{r})\rho_{\rm B}(\boldsymbol{r},[\alpha]), \tag{3.32}$$

where $M$ and $L$ correspond to the coefficients of the linear transformations. Fourier transformations are achieved by setting

$$M(\boldsymbol{k},\boldsymbol{x}) = \exp(-i\boldsymbol{k}\cdot\boldsymbol{x}),\ L(\boldsymbol{p},\boldsymbol{r}) = \exp(-i\boldsymbol{p}\cdot\boldsymbol{r}), \tag{3.33}$$

but we use general notations for a while. Because of the property (2.20), (2.21) of the generalized Wiener-Hermite functionals, the form of equation (3.5) is unchanged under linear transformations (3.32), if we demand the following transformation properties:

$$\psi'^{(n)}(\boldsymbol{k}_1,\ldots,\boldsymbol{k}_n) = \int d^3x_1\cdots d^3x_n M(\boldsymbol{k}_1,\boldsymbol{x}_1)\cdots M(\boldsymbol{k}_n,\boldsymbol{x}_n)\psi^{(n)}(\boldsymbol{x}_1,\ldots,\boldsymbol{x}_n), \tag{3.34}$$

$$K'^{(m)}(\boldsymbol{p};\boldsymbol{k}^{(1)},\ldots,\boldsymbol{k}^{(m)}) = \int d^3r \int d^3x^{(1)}\cdots d^3x^{(m)} L(\boldsymbol{p},\boldsymbol{r}) \times M^{-1}(\boldsymbol{x}^{(1)},\boldsymbol{k}^{(1)})\cdots M^{-1}(\boldsymbol{x}^{(m)},\boldsymbol{k}^{(m)}) K^{(m)}(\boldsymbol{r}-\boldsymbol{x}^{(1)},\ldots,\boldsymbol{r}-\boldsymbol{x}^{(m)}). \tag{3.35}$$

The equation (3.35) has the another expression,

$$K'^{(m)}(\boldsymbol{p};\boldsymbol{k}^{(1)},\boldsymbol{k}^{(m)}) = \left\langle \frac{\delta^m \rho'_{\rm B}(\boldsymbol{p},[\alpha'])}{\delta\alpha'(\boldsymbol{k}^{(1)})\cdots\delta\alpha'(\boldsymbol{k}^{(m)})} \right\rangle_{\rm G}, \tag{3.36}$$

because of equation (3.3). These transformed quantities $\psi'$ and $K'^{(m)}$ are equally applied to the diagrammatic rules i) $\sim$ vi) instead of the original quantities. This gives a way to calculate the correlations of a biased field expressed by linearly transformed quantities (3.32).

Now we restrict the argument to the Fourier transformations (3.33). In this case, from the inverse Fourier transformation, $M^{-1}(\boldsymbol{x},\boldsymbol{k}) = \exp(i\boldsymbol{k}\cdot\boldsymbol{x})/(2\pi)^3$. Then transformed quantities are given by

$$\psi'^{(n)}(\boldsymbol{k}_1,\ldots,\boldsymbol{k}_n) = \tilde{\psi}^{(n)}(\boldsymbol{k}_1,\ldots,\boldsymbol{k}_n) = (2\pi)^3\delta^3(\boldsymbol{k}_1+\cdots+\boldsymbol{k}_n)p^{(n)}(\boldsymbol{k}_1,\ldots,\boldsymbol{k}_{n-1}), \tag{3.37}$$

and

$$K'^{(m)}(\boldsymbol{p};\boldsymbol{k}^{(1)},\ldots,\boldsymbol{k}^{(m)}) = (2\pi)^3\delta^3(\boldsymbol{k}^{(1)}+\cdots+\boldsymbol{k}^{(m)}-\boldsymbol{p})\frac{1}{(2\pi)^{3m}}\widetilde{K}^{(m)}(\boldsymbol{k}^{(1)},\ldots,\boldsymbol{k}^{(m)}), \tag{3.38}$$

where $\widetilde{\psi}^{(n)}$ and $\widetilde{K}^{(m)}$ are the Fourier transformations of $\psi^{(n)}$ and $K^{(m)}$, respectively. Because of the homogeneity of a field $\alpha$, the last expression of equation (3.37) is possible and gives a definition of the multi-spectrum $p^{(n)}$ of a field $\alpha$. In the case $n = 2$, $p^{(2)}$ is called power-spectrum of the field. The Fourier transform of $K^{(m)}$ is given by

$$\widetilde{K}^{(m)}(\boldsymbol{k}^{(1)}, \ldots, \boldsymbol{k}^{(m)}) = (2\pi)^{3m} \left\langle \frac{\delta^m}{\delta\tilde{\alpha}(\boldsymbol{k}^{(1)}) \cdots \delta\tilde{\alpha}(\boldsymbol{k}^{(m)})} \int \frac{d^3p}{(2\pi)^3} \tilde{\rho}_{\rm B}(\boldsymbol{p}, [\tilde{\alpha}]) \right\rangle_{\rm G}, \tag{3.39}$$

from the equations (3.36) and (3.38). This form is convenient if the biasing functional is expressed in Fourier space, noting that the Gaussian averaging in equation (3.39) is given by

$$\langle \cdots \rangle_{\rm G} = \int \cdots \prod_{\boldsymbol{k} \in {\rm uhs}} \left[ \exp\left( -\frac{|\tilde{\alpha}(\boldsymbol{k})|^2}{Vp(k)} \right) \frac{2|\tilde{\alpha}(\boldsymbol{k})| d|\tilde{\alpha}(\boldsymbol{k})|}{Vp(k)} \frac{d\theta_{\boldsymbol{k}}}{2\pi} \right], \tag{3.40}$$

in Fourier space, where uhs means a upper half $\boldsymbol{k}$-space, $\theta_{\boldsymbol{k}}$ is a phase of $\tilde{\alpha}(\boldsymbol{k})$ so that $\tilde{\alpha}(\boldsymbol{k}) = |\tilde{\alpha}(\boldsymbol{k})| \exp(i\theta_{\boldsymbol{k}})$. The power spectrum $p(k)$ is defined by $V^{-1}\langle |\tilde{\alpha}(\boldsymbol{k})|^2 \rangle$, where $V$ is the total volume of the space. Substituting equations (3.37) and (3.38) to our rules, moments $\widetilde{P}_N$ or correlations $\tilde{\xi}^{(N)}_{\rm B}$ in Fourier space are obtained. The multi-spectrum $p^{(N)}_{\rm B}(\boldsymbol{p}_1, \ldots, \boldsymbol{p}_{N-1})$ which is defined by

$$\tilde{\xi}^{(N)}_{\rm B}(\boldsymbol{p}_1, \ldots, \boldsymbol{p}_N) = (2\pi)^3 \delta^3(\boldsymbol{p}_1 + \cdots + \boldsymbol{p}_N) p^{(N)}_{\rm B}(\boldsymbol{p}_1, \ldots, \boldsymbol{p}_{N-1}), \tag{3.41}$$

is more popular than $\tilde{\xi}^{(N)}_{\rm B}$.

The procedures to obtain $\widetilde{P}_N$ are summarized as the following rules:

D-i) Draw $N$ open circles corresponding to $\boldsymbol{p}_1, \ldots, \boldsymbol{p}_N$ which we call external points.

D-ii) Same as ii)

D-iii) Define orientations to all lines and associate labels like $\boldsymbol{k}, \boldsymbol{l}$, etc. to all lines. The orientations of a line $\boldsymbol{k}$ is regarded as a direction of a current $\boldsymbol{k}$. The sum of currents flowing into a vertex should vanish and the sum of currents flowing from a external point $i$ should be $\boldsymbol{p}_i$ (these rules are due to the $\delta$-function in equations (3.37) and (3.38)).

D-iv) Apply the correspondence of Figure 4 for the external points, lines between external points and vertices. Make the product of these factors of the graph.

D-v) Perform integration $\int d^3k/(2\pi)^3$ for each label $\boldsymbol{k}$ and multiply the statistical factor (3.6).

D-vi) Sum up values obtained from all the possible graphs and multiply $(2\pi)^3\delta^3(\boldsymbol{p}_1 + \cdots + \boldsymbol{p}_N)$.

From the value of $\tilde{P}_1$ which can be evaluated by the above rules, the mean density of biased field $P_1 = \bar{\rho}_{\rm B}$ is given by

$$\tilde{P}_1(\boldsymbol{p}) = (2\pi)^3\delta^3(\boldsymbol{p})\bar{\rho}_{\rm B}. \tag{3.42}$$

Using $\bar{\rho}_{\rm B}$, the rules for obtaining the multi-spectrum $p_{\rm B}^{(N)}$ of biased fields are as follows:

E-i) Draw $N$ open circles corresponding to $\boldsymbol{p}_1, \ldots, \boldsymbol{p}_{N-1}$ and $\boldsymbol{p}_N = -(\boldsymbol{p}_1 + \cdots + \boldsymbol{p}_{N-1})$ which we call external points.

E-ii) Same as ii)

E-iii) Define orientations to all lines and associate labels like $\boldsymbol{k}, \boldsymbol{l}$, etc. to all lines. The sum of currents flowing in a vertex should vanish and the sum of currents flowing from a external point $i$ should be $\boldsymbol{p}_i$.

E-iv) Apply the correspondence of Figure 4 for the external points, lines between external points and vertices. Make the product of these factors of the graph.

E-v) Same as D-v)

E-vi) Sum up values obtained from all the connected graphs and multiply $(\bar{\rho}_{\rm B})^{-N}$.

## 4 THEORETICAL APPLICATIONS

### 4.1 The Edgeworth expansion of density probability distribution function

Recently, the Edgeworth expansion of probability distribution function $P(\delta)$ of cosmological density contrast $\delta$ is suggested to be useful in cosmology (Scherrer & Bertschinger 1991; Juszkiewicz et al. 1994; Bernardeau & Kofman 1994). This expansion is suitable to obtain approximately density probability distribution function when restricted number of cumulants are known. The expansion is (Bernardeau & Kofman 1994)

$$\begin{aligned} P(\delta) = \frac{1}{\sqrt{2\pi\sigma^2}} e^{-\nu^2/2} \Bigg[ 1 &+ \sigma\frac{S_3}{6}H_3(\nu) + \sigma^2\left(\frac{S_4}{24}H_4(\nu) + \frac{{S_3}^2}{72}H_6(\nu)\right) \\ &+\sigma^3\left(\frac{S_5}{120}H_5(\nu) + \frac{S_3S_4}{144}H_7(\nu) + \frac{{S_3}^3}{1296}H_9(\nu)\right) + \cdots \Bigg], \end{aligned} \tag{4.1}$$

where $\sigma = \sqrt{\langle\delta^2\rangle}$ is *rms* of density contrast, $\nu = \delta/\sigma$ is a normalized density contrast, and $S_n = \langle\delta^n\rangle_{\rm c}/\langle\delta^2\rangle^{n-1}$ are normalized cumulants. We can see the coefficients of Edgeworth expansion at arbitrary order are very easily calculated in our diagrammatic methods as follows.

Density probability distribution function expressed as

$$P(\delta) = \langle \delta_D(\delta - \nu\sigma) \rangle \tag{4.2}$$

is appropriate to our purpose. Setting $\rho_B = \delta_D(\alpha\sigma - \nu\sigma)$, equation (3.12) reduces to

$$R_m = \frac{1}{\sqrt{2\pi\sigma^2}} e^{-\nu^2/2} H_m(\nu). \tag{4.3}$$

Because $\psi^{(n)}(\boldsymbol{r}, \ldots, \boldsymbol{r}) = S_n \sigma^{n-2}$, a vertex with $n$-lines corresponds to order $\sigma^{n-2}$. Thus, diagrams of Figure 5 (a), (b), (c), (d), ... give all the coefficients of order $\sigma^0$, $\sigma^1$, $\sigma^2$, $\sigma^3$, ..., respectively. Applying rules B-i) $\sim$ B-iv), the expansion (4.1) is easily obtained. This example shows that our diagrammatic method can save the labor of calculation.

### 4.2 Density peaks of Gaussian random fields

We focus on the statistics of density peaks in this subsection. Bardeen et al. (1986) investigated the statistics of density peaks of random Gaussian fields identifying the density peaks of primordial fluctuation with the sites for galaxy or cluster formation. They gave practical approximations to the $N$-point correlation functions of the density peaks. We show, in the following, that our method applied to density peaks of random Gaussian field provides the improved approximations to the correlation functions along the strategy of theirs.

In this subsection, the field $\alpha$ is identified with normalized density fluctuation $\delta_R(\boldsymbol{x})/\langle \delta_R^2 \rangle^{1/2}$. The number density of peaks of the field $\alpha$ greater than a threshold $\nu$ is given by (Bardeen et al. 1986)

$$\rho_{\rm pk}(\boldsymbol{r}) = \theta(\alpha(\boldsymbol{r}) - \nu)\delta^3\left(\partial_i \alpha(\boldsymbol{r})\right)(-1)^3 \det\left(\partial_i\partial_j \alpha(\boldsymbol{r})\right) \theta\left(\lambda_1(\boldsymbol{r})\right) \theta\left(\lambda_2(\boldsymbol{r})\right) \theta\left(\lambda_3(\boldsymbol{r})\right), \tag{4.4}$$

which is a semi-nonlocal bias in our context. In the above, $\lambda_1$, $\lambda_2$, $\lambda_3$ are eigenvalues of a matrix $-(\partial_i\partial_j\alpha)$. The approximation, which is common to Bardeen et al. (1986), is to neglect all the derivatives of the two-point correlation $\psi^{(2)}$. Other correlations $\psi^{(N)}$ ($N \geq 3$) do not exist because of the assumption of Gaussianity of underlying fluctuation. As an illustration, let us consider the case the power spectrum of background fluctuation has the power-law form with spectral index $n$. In this case, the two-point correlation $\psi^{(2)}(\boldsymbol{r})$ falls off as $|\boldsymbol{r}|^{-(n+3)}$ and the $m$-th derivative of the function falls off as $|\boldsymbol{r}|^{-(n+3+m)}$. When $n < -2$, all the derivatives of two-point correlation function can be neglected even if $\psi \sim 1$.

Adopting the above approximation, the rules C-i) $\sim$ C-vi) for the semi-nonlocal bias of random Gaussian underlying field are reduced to the rules A-i) $\sim$ A-vi) for the local bias of random Gaussian field with

$$R_m = \left\langle \left( \frac{\partial}{\partial \alpha_0} \right)^m \rho_{\rm B}(\alpha_\mu) \right\rangle_{\rm G}. \tag{4.5}$$

Note that this simplicity does not hold in non-Gaussian underlying field. Further assumptions for higher-order correlations which can hardly be justified are needed to have this simplicity for general non-Gaussian field (see the next section).

For the density peaks, equation (4.5) can be calculated using equation (A.18) of Bardeen et al. (1986) and results in

$$R_m^{(\mathrm{pk})}(\nu) = \int d\alpha \mathcal{N}_{\mathrm{pk}}^{(\mathrm{G})}(\alpha) \left(\frac{\partial}{\partial \alpha}\right)^m \theta(\alpha - \nu) = \begin{cases} \mathrm{n}_{\mathrm{pk}}^{(\mathrm{G})}(\nu) & (m=0), \\ \left(-\frac{d}{d\nu}\right)^{m-1} \mathcal{N}_{\mathrm{pk}}^{(\mathrm{G})}(\nu) & (m \geq 1). \end{cases} \tag{4.6}$$

where

$$\mathrm{n}_{\mathrm{pk}}^{(\mathrm{G})}(\nu) = \int_\nu^\infty d\alpha \mathcal{N}_{\mathrm{pk}}^{(\mathrm{G})}(\alpha) \tag{4.7}$$

$$\mathcal{N}_{\mathrm{pk}}^{(\mathrm{G})}(\nu) = \frac{1}{(2\pi)^2 R_*^3} e^{-\nu^2/2} G(\gamma, \gamma\nu) \tag{4.8}$$

$$G(\gamma, x_*) = \int_0^\infty dx f(x) \frac{\exp\left[-\frac{(x-x_*)^2}{2(1-\gamma^2)}\right]}{[2\pi(1-\gamma^2)]^{1/2}} \tag{4.9}$$

$$f(x) = \frac{x^3 - 3x}{2}\left\{\mathrm{erf}\left[\left(\frac{5}{2}\right)^{1/2} x\right] + \mathrm{erf}\left[\left(\frac{5}{2}\right)^{1/2} \frac{x}{2}\right]\right\} + \left(\frac{2}{5\pi}\right)^{1/2}\left[\left(\frac{31x^2}{4} + \frac{8}{5}\right) e^{-5x^2/8} + \left(\frac{x^2}{2} - \frac{8}{5}\right) e^{-5x^2/2}\right] \tag{4.10}$$

and erf$(x)$ is an error function:

$$\mathrm{erf}(x) = \frac{2}{\sqrt{\pi}} \int_0^x e^{-x^2} dt. \tag{4.11}$$

Spectral parameters $\sigma_j$, $\gamma$, $R_*$ are defined by

$$\sigma_j^2 = \int \frac{dk}{2\pi^2} k^{2j+2} P(k), \tag{4.12}$$

$$\gamma = \frac{\sigma_1^2}{\sigma_2 \sigma_0}, \quad R_* = \frac{\sqrt{3}\sigma_1}{\sigma_2}, \tag{4.13}$$

where $P(k)$ is a power spectrum:

$$P(k) = \sigma^2 \int d^3x e^{-i\boldsymbol{k}\cdot\boldsymbol{x}} \psi^{(2)}(\boldsymbol{x}). \tag{4.14}$$

equation (4.6) can be represented as one-dimensional integrations as derived in Appendix C:

$$R_m^{(\mathrm{pk})}(\nu) =$$

$$\begin{cases} \dfrac{1}{8\pi^2 R_*^3}\displaystyle\int_0^\infty dx f(x) e^{-x^2/2} \mathrm{erfc}\left(\frac{\nu - \gamma x}{\sqrt{2(1-\gamma^2)}}\right) & (m=0) \\ \dfrac{e^{-\nu^2/2}}{(2\pi)^{5/2} R_*^3 (1-\gamma^2)^{m/2}} \displaystyle\int_0^\infty dx f(x) H_{m-1}\left(\frac{\nu-\gamma x}{\sqrt{1-\gamma}}\right) \exp\left[-\frac{(x-\gamma\nu)^2}{2(1-\gamma^2)}\right] & (m \geq 1) \end{cases} \tag{4.15}$$

which have to be integrated numerically. In the high $\nu$ limit, these coefficients reduce to

$$R_m^{(\mathrm{pk})}(\nu) \stackrel{\nu\to\infty}{\longrightarrow} \frac{1}{12\sqrt{3}\pi^2} \frac{\sigma_1}{\sigma_0} H_{m+2}(\nu) e^{-\nu^2/2}. \tag{4.16}$$

To see the behavior of two-point correlation function of density peaks on large scales which was described in Bardeen et al. (1986), we consider the case, $\psi^{(2)} \to 0$. The two-point correlation function of peaks calculated by our method is, to lowest order in $\psi^{(2)}$,

$$\xi_{\mathrm{pk}}(\boldsymbol{r}_1, \boldsymbol{r}_2) = \left(\frac{\mathcal{N}_{\mathrm{pk}}^{(\mathrm{G})}(\nu)}{\mathrm{n}_{\mathrm{pk}}^{(\mathrm{G})}(\nu)}\right)^2 \psi^{(2)}(\boldsymbol{r}_1, \boldsymbol{r}_2). \tag{4.17}$$

The corresponding expression in Bardeen et al. (1986) [equation (6.12) of their paper] is

$$\xi_{\mathrm{pk}} = \langle\tilde{\alpha}\rangle^2 \psi^{(2)}, \tag{4.18}$$

where

$$\langle\tilde{\alpha}\rangle \equiv \frac{1}{(2\pi)^2 R_*^3 \mathrm{n}_{\mathrm{pk}}^{(\mathrm{G})}(\nu)} \int_\nu^\infty d\alpha e^{-\alpha^2/2} \int_0^\infty dx f(x) \frac{\exp\left[-\dfrac{(x-\gamma\alpha)^2}{2(1-\gamma^2)}\right]}{[2\pi(1-\gamma^2)]^{1/2}} \frac{\alpha - \gamma x}{1-\gamma^2}. \tag{4.19}$$

The above two expressions are equivalent. In fact, explicit calculation (see Appendix D) shows

$$\langle\tilde{\alpha}\rangle = \frac{\mathcal{N}_{\mathrm{pk}}^{(\mathrm{G})}(\nu)}{\mathrm{n}_{\mathrm{pk}}^{(\mathrm{G})}(\nu)}. \tag{4.20}$$

In high $\nu$ limit, $\xi_{\mathrm{pk}} = \nu^2\psi^{(2)}$ which corresponds to Kaiser's result (Kaiser 1984).

Jensen and Szalay (1986) give the expression of the autocorrelation functions of regions in which the value of a field $\alpha$ is higher than a threshold $\nu$. Autocorrelation functions of the thresholded regions are regarded as approximations for correlation functions of peaks in high $\nu$ limit in the context of biased structure formation in astrophysics. Their formula can be derived by our method as shown in the previous section: equations (3.19) and (3.22) derive their formula [eq. (4) of their paper],

$$\frac{P_N(\boldsymbol{r}_1, \ldots, \boldsymbol{r}_N)}{P_1^N} = \sum_{m=0}^\infty \sum_{\{m_{kl}|m\}} \prod_{k<l} \frac{[\psi(\boldsymbol{r}_k, \boldsymbol{r}_l)]^{m_{kl}}}{m_{kl}!} \cdot \prod_{r=1}^N A_{m_r}^{(\mathrm{sc})}, \tag{4.21}$$

where

$$A_m^{(\rm sc)} = \frac{R_m^{(\rm sc)}}{R_0^{(\rm sc)}} = \begin{cases} 1 & (m=0), \\ \dfrac{\sqrt{2}H_{m-1}(\nu)}{\sqrt{\pi}e^{\nu^2/2}{\rm erfc}(\nu/\sqrt{2})} & (m \geq 1). \end{cases} \tag{4.22}$$

Note that the definition of the Hermite polynomials are different from that in Jensen & Szalay (1986). In the case of density peaks, it is obvious that the above quantities $A_m^{(\rm sc)}$ should be replaced as follows:

$$A_m^{(\rm sc)}(\nu) \rightarrow$$

$$A_m^{(\rm pk)} = \begin{cases} 1 & (m=0), \\ \sqrt{\dfrac{2}{\pi}} \dfrac{e^{-\nu^2/2}}{(1-\gamma^2)^{m/2}} \dfrac{\displaystyle\int_0^\infty dx f(x) H_{m-1}\left(\frac{\nu-\gamma x}{\sqrt{1-\gamma^2}}\right) \exp\left[-\frac{(x-\gamma\nu)^2}{2(1-\gamma^2)}\right]}{\displaystyle\int_0^\infty dx f(x) e^{-x^2/2} {\rm erfc}\left(\frac{\nu-\gamma x}{\sqrt{2(1-\gamma^2)}}\right)} & (m \geq 1). \end{cases} \tag{4.23}$$

This quantities are easily evaluated by one-dimensional numerical integrations. This expression for $N$-point correlation function [eq. (4.21) and eq. (4.23)] of density peaks in Gaussian random fields is efficient than the methods developed in Bardeen et al. (1986), especially for higher $N$. In practice, the expansion is summed for arbitrary accuracy as in the Jensen and Szalay's formula. Figure 6 shows the various approximations for the two-point correlation function of peaks including the approximations of Jensen and Szalay [eq. (6) of Jensen & Szalay (1986)], Bardeen et al. [eq. (6.22) of their paper with $n=2$], peak-background splitting method developed by Bardeen et al. [eq. (6.46) of their paper], Lumsden et al. (1989)$^\dagger$, comparing with our method [eq. (4.21) and eq. (4.23)]. In the plot, the underlying fluctuation is assumed to be random Gaussian. We use the smoothed power spectrum of standard cold dark matter model given in Bardeen et al. (1986) with parameters $\Omega = 1, \Lambda = 0$, and a Hubble constant of 50 km/s/Mpc. The smoothing length $R$ is $0.2h^{-1}$Mpc and the Gaussian window (2.2) is adopted. Figure 7 shows the three-point correlation functions of peaks in equilateral configuration for the approximations of Jensen and Szalay [eq. (4) of their paper with $N=3$], peak-background splitting method [eq. (6.53) of Bardeen et al. (1986)], Jensen and Szalay with effective thresholds, comparing with our method for the same underlying spectrum as in Figure 6. In the effective threshold method, we use the same effective threshold as in the two-point correlation function. In Figure 8 plotted the normalized three-point correlation function

$$Q \equiv \frac{\xi^{(3)}(\boldsymbol{r}_1,\boldsymbol{r}_2,\boldsymbol{r}_3)}{\xi^{(2)}(\boldsymbol{r}_1,\boldsymbol{r}_2)\xi^{(2)}(\boldsymbol{r}_2,\boldsymbol{r}_3) + \xi^{(2)}(\boldsymbol{r}_2,\boldsymbol{r}_3)\xi^{(2)}(\boldsymbol{r}_3,\boldsymbol{r}_1) + \xi^{(2)}(\boldsymbol{r}_3,\boldsymbol{r}_1)\xi^{(2)}(\boldsymbol{r}_1,\boldsymbol{r}_2)}, \tag{4.24}$$

for approximations same as in Figure 7.

---

$^\dagger$They used Jensen and Szalay's formula with effective threshold $\nu_{\rm eff}$ which is determined by matching the two-point correlation function of Jensen and Szalay with that of peaks in large-separation limit, i.e., $\nu_{\rm eff}$ is the solution of the implicit equation, $A_1^{(\rm sc)}(\nu_{\rm eff}) = \langle\tilde{\alpha}\rangle(\nu)$ for each true threshold $\nu$.

### 4.3 Weighted density extrema of non-Gaussian random field

The approximation adopted in the previous subsection does not work when the background fluctuation is not Gaussian random field. This is because in calculating the biasing of the non-Gaussian random field, the higher order correlations at the same point including derivatives such as $\langle\alpha^2(\boldsymbol{x})\triangle\alpha(\boldsymbol{x})\rangle$ can not reasonably neglected. In this subsection, we consider the density extrema with weighted factor according to the second derivatives of the field (see below) above some threshold $\nu$, which can serve as an approximation of the density peaks for high threshold. The method to evaluate the density extrema of non-Gaussian random field to arbitrary accuracy is given in the following.

The number density of weighted density extrema above threshold $\nu$ we consider in this subsection is

$$\rho_{\rm ext}(\boldsymbol{r}) = \theta(\alpha(\boldsymbol{r}) - \nu)\delta^3\left(\partial_i\alpha(\boldsymbol{r})\right)(-1)^3\det\left(\partial_i\partial_j\alpha(\boldsymbol{r})\right), \tag{4.25}$$

[*cf.* eq. (4.4)]. This number density field is for density extrema with weight factor $-{\rm sign}[\det(\partial_i\partial_j\alpha(\boldsymbol{r}))]$. For large threshold $\nu$, the number of maxima dominates the number of minima and saddle points, and equation (4.25) is a reasonable approximation for density peaks because the weight factor for density maxima is $+1$. Otto et al. (1986) and Cline et al. (1987) investigated the average number density and the correlation functions of this type of weighted extrema for Gaussian random fields approximately. We are in the place to generalize these analyses for non-Gaussian random fields.

The coefficients for the semi-nonlocal bias (4.25) is

$$\begin{aligned}
R_{\mu_1\cdots\mu_m} &= R(k; l_1, l_2, l_3; p_{11}, p_{22}, p_{33}, p_{23}, p_{13}, p_{12}) \\
&\equiv \frac{1}{(2\pi)^2}\left(\frac{\sigma_1}{\sqrt{3}\sigma_0}\right)^{3-\sum_i l_i - 2\sum_{i\leq j} p_{ij}} H_{l_1}(0)H_{l_2}(0)H_{l_3}(0) \\
&\qquad\times e^{-\nu^2/2}\left[H_{k+2}(\nu)J_0(\{p_{ij}\}) - H_{k+1}(\nu)J_1(\{p_{ij}\})\right. \\
&\qquad\qquad\left.+H_k(\nu)J_2(\{p_{ij}\}) - H_{k-1}(\nu)J_3(\{p_{ij}\})\right].
\end{aligned} \tag{4.26}$$

The derivation of this result is given in Appendix E. In the above notation, the number of indices corresponding to $\alpha$, $\partial\alpha/\partial x^i$, $\partial^2\alpha/\partial x^i\partial x^j$ $(i \leq j)$ among $\mu_1, \ldots, \mu_m$ are $k$, $l_i$, $p_{ij}$, respectively. For $k = 0$, we use the notation,

$$H_{-1}(\nu) \equiv \sqrt{\frac{\pi}{2}}e^{\nu^2/2}{\rm erfc}\left(\frac{\nu}{\sqrt{2}}\right). \tag{4.27}$$

Table I gives the definition of $J_0$, $J_1$, $J_2$ and $J_3$ as functions of $\{p_{ij}\}$. For cases not listed in Table I, $J_i$'s are all vanish. We can calculate the number density and the correlation functions of weighted density extrema of non-Gaussian field to arbitrary accuracy in principle by using the coefficients (4.26) and rules C-i) $\sim$ C-iv) provided that the diagrammatic expansion converges. As primary examples, the number density of weighted extrema $n_{\rm ex}(\nu)$ and the two-point correlation function $\xi_{{\rm ex},\nu}(r)$ of Gaussian random fields to leading order in $\psi$ are, from diagrams in Figure 9,

$$n_{\rm ex}(\nu) = \frac{1}{(2\pi)^2}\left(\frac{\sigma_1}{\sqrt{3}\sigma_0}\right)^3 e^{-\nu^2/2} H_2(\nu), \tag{4.28}$$

$$\xi_{{\rm ex},\nu}(r) = \nu^2\left(\frac{\nu^2-3}{\nu^2-1}\right)^2 \frac{\xi(r)}{\sigma_0^2} - \frac{6\nu^2(\nu^2-3)}{(\nu^2-1)^2}\frac{\triangle\xi(r)}{\sigma_1^2} + \frac{9\nu^2}{(\nu^2-1)^2}\frac{\sigma_0^2\triangle^2\xi(r)}{\sigma_1^4}, \tag{4.29}$$

where $\xi(r) = \sigma_0^2\psi(r)$. Otto et al. (1986) and Catelan et al. (1988) derived the leading and sub-leading contribution of $\nu$ to $\xi_{{\rm ex},\nu}$. Our method provides a way to calculate the same quantity not relying on the expansion in $\nu$.

In the high $\nu$ limit, the leading contribution of $\nu$ in each diagram is come from the term which maximize $k$ for each external point as seen from the expression (4.26). Thus, considering the leading contribution of $\nu$ is equivalent to considering the local bias with coefficients

$$R_k = \frac{1}{(2\pi)^2}\left(\frac{\sigma_1}{\sqrt{3}\sigma_0}\right)^3 \nu^{k+2} e^{-\nu^2/2}. \tag{4.30}$$

Substituting these coefficients to equation (3.17), we obtain

$$P_N = \left[\frac{1}{(2\pi)^2}\left(\frac{\sigma_1}{\sqrt{3}\sigma_0}\right)^3 \nu^2 e^{-\nu^2}\right]^N Z_N(J_i = \nu). \tag{4.31}$$

The number density is

$$P_1 = \frac{1}{(2\pi)^2}\left(\frac{\sigma_1}{\sqrt{3}\sigma_0}\right)^3 \nu^2 e^{-\nu^2} Z_1(\nu), \tag{4.32}$$

which was previously derived by Catelan et al. (1988) by using a different method. The correlation function is calculated by taking the connected part of the following quantity,

$$\frac{P_N}{P_1^N} = \frac{Z_N(J_i=\nu)}{[Z_1(\nu)]^N}, \tag{4.33}$$

which is the same as in the biasing with sharp clipping in high $\nu$ limit [eq. (3.25)].

### 4.4 Biasing and hierarchical underlying fluctuations

In this subsection, we consider the case that the underlying fluctuation satisfy the hierarchical model of correlation functions (White 1979; Fry 1984);

$$\xi^{(N)}(\boldsymbol{x}_1,\ldots,\boldsymbol{x}_N) = \sum_{\text{trees }(a)} Q_a^{(N)} \sum_{\text{labelings}} \prod_{\text{edges }(AB)}^{N-1} \xi^{(2)}(\boldsymbol{x}_A,\boldsymbol{x}_B). \tag{4.34}$$

In the above symbolic notation, the edge $(AB)$ is one of the edges in a tree graph $(a)$ which is a set of connected $N-1$ edges linking $N$ points, $\boldsymbol{x}_1, \ldots, \boldsymbol{x}_N$ without making any loop. All the distinct tree graphs are labeled by $(a)$ and "labelings" indicate the symmetric sum with respect to the $N$-points (Fry 1984). $Q_a^{(N)}$ are constants for each tree topology $(a)$. If $Q_a^{(N)} = 0$ for all $N \geq 3$, the hierarchical model reduces to random Gaussian fluctuation. Fry & Gaztañaga (1993) showed that the local bias which can be expanded as Taylor series by density contrast $\delta$ of background field,

$$\rho_{\mathrm{B}}(\delta) = \sum_{k=0}^{\infty} \frac{a'_k}{k!} \delta^k, \tag{4.35}$$

leads the hierarchical form (4.34) of biased field in the leading order of $\xi^{(2)}$. In this argument, the coefficients $a'_k$ are assumed to be the zero-th order of $\xi^{(2)}$. This argument can not be applied to biasing by sharp clipping (3.21) because sharp clipping can not be expanded as Taylor series. Moreover, even if we approximate the sharp clipping by some smooth function, it is a function of $\delta/\sigma$ and the coefficients $a'_k$ are no longer the zero-th order of $\xi^{(2)} \sim \mathcal{O}(\sigma^2)$. We complement, in the following, Fry and Gaztañaga's argument by investigate the case biasing is a local function of normalized density contrast $\delta/\sigma$ which has not to be expanded as a Taylor series as in the sharp clipping bias or in density peaks in the approximation of the preceding subsection.

In the following, the background field $\alpha$ is identified with the normalized density contrast $\delta/\sigma$. The field $\delta$ is also assumed to satisfy the hierarchical model (4.34). The $j$-vertex, in this case, contributes the order $\sigma^{j-1}$ and we need not to consider any vertex if we only consider the leading order in $\sigma$. Evaluating diagrams in Figure 10, the following results are obtained to the leading order both in $\sigma$ and $\psi^{(2)}$:

$$P_1 = R_0, \tag{4.36}$$

$$\xi^{(2)}_{\mathrm{B},12} = \left(\frac{R_1}{R_0}\right)^2 \psi_{12}, \tag{4.37}$$

$$\xi^{(3)}_{\mathrm{B},123} = \frac{R_1^2 R_2}{R_0^3} (\psi_{12}\psi_{23} + \mathrm{cyc.}), \tag{4.38}$$

$$\xi^{(4)}_{\mathrm{B},1234} = \frac{R_1^2 R_2^2}{R_0^4} (\psi_{12}\psi_{23}\psi_{34} + \mathrm{sim.}(12)) + \frac{R_1^3 R_3}{R_0^4} (\psi_{12}\psi_{13}\psi_{14} + \mathrm{sim.}(4)). \tag{4.39}$$

The higher orders of $\psi^{(2)}$ is omitted for we are interested in large-separation limit. Thus, the parameters $Q_{N,a}^{(B)}$ of hierarchical model of a biased field are

$$Q^{(\mathrm{B})}_{3,\alpha} = \frac{R_0 R_2}{R_1^2}, \tag{4.40}$$

$$Q^{(\mathrm{B})}_{4,\beta} = \frac{R_0^2 R_2^2}{R_1^4}, \qquad Q^{(\mathrm{B})}_{4,\gamma} = \frac{R_0^2 R_3}{R_1^3}, \tag{4.41}$$

to leading order. The indices of tree graphs $\alpha$, $\beta$, $\gamma$ are defined in Figure 11. For arbitrary $N$, it is obvious from above calculation that $\xi_{\mathrm{B}}^{(N)}$ satisfy the hierarchical model to leading order in $\xi_{\rm B}^{(2)} \sim \mathcal{O}(\sigma^2)$ and, moreover, the parameters $Q_{N,a}^{(\rm B)}$ can be directly obtained from the topology $(a)$ by simple rules as follows: let $m_j$ be the number of external points which is attached by $j$ lines in a connected tree graph $(a)$ ($\sum_j m_j = N$, $\sum_j j m_j = 2(N-1)$), then

$$Q_{N,a}^{(\rm B)} = \left(\frac{R_1}{R_0}\right)^{-2(N-1)} \frac{\prod_j R_j^{m_j}}{R_0^N} = A_1^{-2(N-1)} \prod_j A_j^{m_j}, \tag{4.42}$$

where $A_j = R_j/R_0$. In the biasing by sharp clipping, equation (4.22) is applied. In the biasing by density peaks, equation (4.23) is applied in the approximation of the preceding subsection. In high $\nu$ limit, both biasing via sharp clipping and via density peaks have the same behavior, $A_m(\nu) \stackrel{\nu\to\infty}{\longrightarrow} \nu^m$, and always $Q_{N,a}^{(\rm B)} = 1$.

### 4.5 Gravitational evolution as nonlocal biasing

In our formalism, the gravitational nonlinear evolution of the density field can be seen as a kind of nonlocal biasing from the initial fluctuation field. In this subsection, we give explicit relation between our formalism and the perturbation theory of gravitational instability and apply our formalism to the calculation of correlation functions of density field. We consider the non-relativistic collisionless self-gravitating system in the fluid limit for Einstein-de Sitter universe ($\Omega = 1$, $\Lambda = 0$, $p = 0$) as an example. The mass density contrast $\delta(\boldsymbol{x}, t)$ and the peculiar velocity field $\boldsymbol{v}(\boldsymbol{x}, t)$ are governed by the equations (Peebles 1980):

$$\frac{\partial \delta}{\partial t} + \frac{1}{a}\nabla \cdot [(1+\delta)\boldsymbol{v}] = 0, \tag{4.43}$$

$$\frac{\partial \boldsymbol{v}}{\partial t} + \frac{\dot{a}}{a}\boldsymbol{v} + \frac{1}{a}(\boldsymbol{v}\cdot\nabla)\boldsymbol{v} + \frac{1}{a}\nabla\phi = 0, \tag{4.44}$$

$$\triangle\phi = 4\pi G\bar{\rho}a^2\delta, \tag{4.45}$$

where $\nabla$ denotes the derivative by means of comoving coordinate $\boldsymbol{x}$, $\bar{\rho}$ and $a(t)$ are the average density and expansion factor, respectively, and are given by $\bar{\rho} = (6\pi G t^2)^{-1}$, $a \propto t^{2/3}$ in the Einstein-de Sitter universe. The linear solution is derived by linearizing equations (4.43)-(4.45). The growing mode of the linear solution has the form,

$$\delta^{(1)}(\boldsymbol{x}, t) = \left(\frac{t}{t_{\rm r}}\right)^{2/3} \epsilon(\boldsymbol{x}), \tag{4.46}$$

where $t_{\rm r}$ is initial time or recombination time, $\epsilon(\boldsymbol{x})$ is interpreted as the primordial fluctuation. The second order perturbative solution for growing mode (Peebles 1980) is given by

$$\delta^{(2)}(\boldsymbol{x}, t) = \left(\frac{t}{t_{\rm r}}\right)^{4/3} \left\{\frac{5}{7}\epsilon^2(\boldsymbol{x}) + \nabla\epsilon(\boldsymbol{x})\cdot\nabla U(\boldsymbol{x}) + \frac{2}{7}\left[\partial_i\partial_j U(\boldsymbol{x})\right]\left[\partial_i\partial_j U(\boldsymbol{x})\right]\right\}, \tag{4.47}$$

where

$$U(\boldsymbol{x}) = -\frac{1}{4\pi}\int d^3x' \frac{\epsilon(\boldsymbol{x}')}{|\boldsymbol{x}-\boldsymbol{x}'|}. \tag{4.48}$$

The mass density,

$$\rho_{\rm pt}(\boldsymbol{x},t) = \bar{\rho}(t)[1+\delta^{(1)}(\boldsymbol{x},t)+\delta^{(2)}(\boldsymbol{x},t)+\mathcal{O}(\epsilon^3)], \tag{4.49}$$

is regarded as nonlocal biasing where the background field $\alpha$ is identified with normalized primordial fluctuation $\epsilon/\sigma$, where $\sigma^2 = \langle\epsilon^2\rangle$. The kernels are derived to be

$$K^{(0)} = \bar{\rho}(t), \tag{4.50}$$

$$K^{(1)}(\boldsymbol{x}) = \sigma\left(\frac{t}{t_{\rm r}}\right)^{2/3}\bar{\rho}(t)\delta^3(\boldsymbol{x}) + \mathcal{O}(\sigma^3), \tag{4.51}$$

$$K^{(2)}(\boldsymbol{x},\boldsymbol{y}) = \sigma^2\left(\frac{t}{t_{\rm r}}\right)^{4/3}\bar{\rho}(t)\left[\frac{10}{7}\delta^3(\boldsymbol{x})\delta^3(\boldsymbol{y}) - \frac{1}{4\pi}\frac{\partial\delta^3(\boldsymbol{x})}{\partial x_i}\frac{\partial|\boldsymbol{y}|^{-1}}{\partial y_i} - \frac{1}{4\pi}\frac{\partial\delta^3(\boldsymbol{y})}{\partial y_i}\frac{\partial|\boldsymbol{x}|^{-1}}{\partial x_i} + \frac{1}{28\pi^2}\frac{\partial^2|\boldsymbol{x}|^{-1}}{\partial x_i\partial x_j}\frac{\partial^2|\boldsymbol{y}|^{-1}}{\partial y_i\partial y_j}\right] + \mathcal{O}(\sigma^4), \tag{4.52}$$

$$K^{(n)} = \mathcal{O}(\sigma^n) \qquad (n\geq 3), \tag{4.53}$$

which are nonlocal. Using these kernels, and assuming that the primordial fluctuation $\epsilon$ is Gaussian, our rules A-i) $\sim$ A-vi) for two- and three-point correlation function give the following results after lengthy algebra:

$$\xi_{\rm pt}(\boldsymbol{r}_1,\boldsymbol{r}_2;t) = \left(\frac{t}{t_{\rm r}}\right)^{4/3}\xi_{\rm in}(r_{12}) + \mathcal{O}(\sigma^4), \tag{4.54}$$

$$\begin{aligned}\zeta_{\rm pt}(\boldsymbol{r}_1,\boldsymbol{r}_2,\boldsymbol{r}_3;t) = \left(\frac{t}{t_{\rm r}}\right)^{8/3}\Bigg[\!\!\Bigg[&\frac{2}{7}(5+2\cos^2\theta_2)\xi_{\rm in}(r_{12})\xi_{\rm in}(r_{23})\\ &+\cos\theta_2\left[\frac{\xi'_{\rm in}(r_{12})J_3^{\rm in}(r_{23})}{r_{23}^2}+\frac{\xi'_{\rm in}(r_{23})J_3^{\rm in}(r_{12})}{r_{12}^2}\right]\\ &+\frac{4}{7}\Bigg\{3(3\cos^2\theta_2-1)\frac{J_3^{\rm in}(r_{12})J_3^{\rm in}(r_{23})}{r_{12}^3r_{23}^3}\\ &+(1-3\cos^2\theta_2)\left[\frac{\xi_{\rm in}(r_{12})J_3^{\rm in}(r_{23})}{r_{23}^3}+\frac{\xi_{\rm in}(r_{23})J_3^{\rm in}(r_{12})}{r_{12}^3}\right]\Bigg\}\Bigg]\!\!\Bigg]\\ &+{\rm cyc.}(3\ {\rm terms}) + \mathcal{O}(\sigma^6).\end{aligned} \tag{4.55}$$

In the above, the two-point correlation function of the primordial field is

$$\xi_{\rm in}(r_{12}) = \langle\epsilon(\boldsymbol{r}_1)\epsilon(\boldsymbol{r}_2)\rangle \tag{4.56}$$

where $r_{12} = |\boldsymbol{r}_1-\boldsymbol{r}_2|$; $\theta_2$ is an angle between two vectors, $\boldsymbol{r}_1-\boldsymbol{r}_2$ and $\boldsymbol{r}_3-\boldsymbol{r}_2$; and $J_3^{\rm in}(r) = \int_0^r dr'^2\xi_{\rm in}(r')$. Recently, Bharadwaj (1994) independently obtained the same result for three-point correlation function (4.55). Putting $\xi_{\rm in} \propto r^{\gamma_{\rm in}}$ in equation (4.55), we rederive Fry's result (Fry 1984). For equilateral configuration, $r_{12} = r_{23} = r_{31}$, the hierarchical amplitude, $Q_{pt} = \zeta_{\rm pt}/(3\xi_{\rm pt}^2)$ is

$$Q_{\rm pt}({\rm equilateral}) = \frac{18\gamma_{\rm in}^2 - 89\gamma_{\rm in} + 102}{7(3-\gamma_{\rm in})^2}. \tag{4.57}$$

## 5 CONCLUSIONS

We have developed a formalism to explore the statistics of biased field in large-scale structure of the universe. In particular, the general nonlocal biasing can be treated in this formalism. The underlying field is not necessarily assumed to be a random Gaussian field. The formalism is based on diagrammatic series expansion of the correlation functions of a biased field. As for local biasing, the form of series expansion of a biased field is known by previous works, but it is very complicated expression (Borgani & Bonometto 1989, 1990). Our formalism, when applied to local biasing problem, simplify this complexity.

In quantum field theory and statistical physics, the Feynman diagrams play roles more than mere simplification of calculations. It would be interesting if our formalism could play similar roles. In fact, our diagrammatic expansion can be conceived to be the generalization of the Edgeworth expansion which is useful in connecting dynamics and statistics in large-scale structure by the argument in §4.1.

We have applied our formalism to the problems which have been studied previously to show the effectiveness of our method. Application to peak statistics improves the technique developed by Bardeen et al. (1986). In particular, we can evaluate the higher order correlations of peaks successfully for which technique of Bardeen *et al.* had difficulty. For statistics of weighted extrema, which could be an approximation to peak statistics, we showed the general way of calculation not depending on the high-threshold expansion. Weighted extrema *is* observable in surveys of large-scale structure and our result can be directly compared with observations. The similar technique was recently applied to the genus statistics of weakly non-Gaussian field by the present author (Matsubara 1994). The conservation of hierarchical model of correlation functions subjected to local biasing, which is studied by Fry and Gaztañaga (1993), was revisited using our formalism. We complement their results by investigating a local biasing through normalized density contrast which may not be represented by Taylor series, as in sharp-clipping biasing. The gravitational evolution of primordial fluctuation also falls under the category of nonlocal biasing as shown in §4.5.

Those problems have been studied individually before this paper. Our formalism enables more detailed examination of those problems. Those examples show that our formalism is not only a general method but also a practically powerful way to investigate various problems. The formalism developed in this paper may be an influential method for future investigations of large-scale structure of the universe.

## ACKNOWLEDGEMENTS

I thank Y. Suto, and J. Yokoyama for helpful discussions. I thank T. Watanabe for his help in plotting Figure 6, 7 and 8 and discussions.

I acknowledge the support of JSPS Fellowship. This research was supported in part by the Grants-in-Aid for Scientific research from Ministry of Education, Science and Culture of Japan (No. 0042).

## APPENDIX A: EXPECTATION VALUES OF PRODUCTS OF GENERALIZED WIENER-HERMITE FUNCTIONALS

Imamura, Meecham & Siegel (1965) gives the method to calculate symbolically the expectation values of products of Wiener-Hermite functionals. In this Appendix, we generalize their method to the case of generalized Wiener-Hermite functionals. The quantity we need to calculate is

$$\left\langle \prod_{i=1}^{s} \mathcal{H}_{(m_i)}\left(\boldsymbol{x}_i^{(1)}, \ldots, \boldsymbol{x}_i^{(m_i)}\right) \right\rangle_{\mathrm{G}}. \tag{A.1}$$

To derive the method to calculate this quantity, there are at least two ways, i.e., using operators and using generating functionals. We present both ways here although either is sufficient to our purpose.

### A1 The method using operators

The method using operators is to use the operator representation of generalized Wiener-Hermite functionals. This method is similar to the derivation of Feynman rules in operator formalism in quantum field theory.

We define the annihilation operator $a(\boldsymbol{x})$ and creation operator $a^\dagger(\boldsymbol{x})$ as

$$a(\boldsymbol{x}) = \frac{1}{2}\int d^3y \psi^{-1}(\boldsymbol{x},\boldsymbol{y})\alpha(\boldsymbol{y}) + \frac{\delta}{\delta\alpha(\boldsymbol{x})}, \tag{A.2}$$

$$a^\dagger(\boldsymbol{x}) = \frac{1}{2}\alpha(\boldsymbol{x}) - \int d^3y \psi(\boldsymbol{x},\boldsymbol{y})\frac{\delta}{\delta\alpha(\boldsymbol{y})}. \tag{A.3}$$

Annihilation and creation operators transform as covariant and contravariant vectors, respectively under the linear transformation (2.19). These operators indeed satisfy the following commutation relations as annihilation and creation operators:

$$\left[a(\boldsymbol{x}), a^\dagger(\boldsymbol{y})\right] = \delta^3(\boldsymbol{x}-\boldsymbol{y}), \tag{A.4}$$

$$\left[a(\boldsymbol{x}), a(\boldsymbol{y})\right] = \left[a^\dagger(\boldsymbol{x}), a^\dagger(\boldsymbol{y})\right] = 0. \tag{A.5}$$

In terms of annihilation and creation operators, the generalized Wiener-Hermite functionals are represented by

$$\mathcal{H}_{(m)}\left(\boldsymbol{x}_1, \ldots, \boldsymbol{x}_m\right) = \,:\prod_{i=1}^{m}\left[\tilde{a}(\boldsymbol{x}_i) + a^\dagger(\boldsymbol{x}_i)\right]:, \tag{A.6}$$

where

$$\tilde{a}(\boldsymbol{x}) \equiv \int d^3y \psi(\boldsymbol{x},\boldsymbol{y}) a(\boldsymbol{y}), \tag{A.7}$$

is a contravariant vector. The notation $:\cdots:$ means the Wick product (Wick 1950): place all the annihilation operators to the left of all the creation operators in the Wick product. For example,

$$:\left[\tilde{a}(\boldsymbol{x})+a^\dagger(\boldsymbol{x})\right]\left[\tilde{a}(\boldsymbol{y})+a^\dagger(\boldsymbol{y})\right]:=\tilde{a}(\boldsymbol{x})\tilde{a}(\boldsymbol{y})+\tilde{a}(\boldsymbol{x})a^\dagger(\boldsymbol{y})+\tilde{a}(\boldsymbol{y})a^\dagger(\boldsymbol{x})+a^\dagger(\boldsymbol{x})a^\dagger(\boldsymbol{y}), \quad \text{(A.8)}$$

and so on. The representation (A.6) is proved by induction using the recursion relation (2.23).

We then define the "vacuum state" $|0\rangle$ as

$$|0\rangle \equiv \{P_{\mathrm{G}}[\alpha]\}^{1/2}, \tag{A.9}$$

where $P_{\mathrm{G}}[\alpha]$ is an infinite dimensional Gaussian distribution functional given by equation (2.24). The "$n$ particle states" ($n \geq 0$) are defined using vacuum state as

$$|\boldsymbol{x}_1,\ldots,\boldsymbol{x}_n\rangle = a^\dagger(\boldsymbol{x}_1)\cdots a^\dagger(\boldsymbol{x}_n)|0\rangle, \tag{A.10}$$

which are the bases of the Fock space $\mathcal{F}$. The following properties hold for $n$ particle state:

$$a^\dagger(\boldsymbol{y})|\boldsymbol{x}_1,\ldots,\boldsymbol{x}_n\rangle = |\boldsymbol{y},\boldsymbol{x}_1,\ldots,\boldsymbol{x}_n\rangle, \tag{A.11}$$

$$a(\boldsymbol{y})|0\rangle = 0, \tag{A.12}$$

$$a(\boldsymbol{y})|\boldsymbol{x}_1,\ldots,\boldsymbol{x}_n\rangle = \sum_{i=1}^{n}\delta^3(\boldsymbol{y}-\boldsymbol{x}_i)|\boldsymbol{x}_1,\ldots,\boldsymbol{x}_{i-1},\boldsymbol{x}_{i+1},\ldots,\boldsymbol{x}_n\rangle. \tag{A.13}$$

We define the inner product of $n$ particle state $|\boldsymbol{x}_1,\ldots,\boldsymbol{x}_n\rangle$ and $m$ particle state $|\boldsymbol{y}_1,\ldots,\boldsymbol{y}_m\rangle$ in the Fock space $\mathcal{F}$ by

$$\begin{aligned}\langle\boldsymbol{x}_1,\ldots,\boldsymbol{x}_n|\boldsymbol{y}_1,\ldots,\boldsymbol{y}_m\rangle \equiv \delta_{nm}\int d^3z_1\cdots d^3z_n\psi^{-1}(\boldsymbol{x}_1,\boldsymbol{z}_1)\cdots\psi^{-1}(\boldsymbol{x}_n,\boldsymbol{z}_n)\\ \times\int[d\alpha]|\boldsymbol{z}_1,\ldots,\boldsymbol{z}_n\rangle|\boldsymbol{y}_1,\ldots,\boldsymbol{y}_n\rangle,\end{aligned} \tag{A.14}$$

where we employ Dirac's notation for the inner product. The annihilation operator and the creation operator are Hermite conjugate to each other under this definition of the inner product in the Fock space. The inner product (A.14) can be calculated explicitly by commutation relations (A.4), (A.5) and the properties (A.13) resulting in

$$\begin{aligned}\langle\boldsymbol{x}_1,\ldots,\boldsymbol{x}_n|\boldsymbol{y}_1,\ldots,\boldsymbol{y}_m\rangle &= \left\langle 0\left|a(\boldsymbol{x}_1)\cdots a(\boldsymbol{x}_n)a^\dagger(\boldsymbol{y}_1)\cdots a^\dagger(\boldsymbol{y}_m)\right|0\right\rangle\\ &= \delta_{mn}\left[\delta^3(\boldsymbol{x}_1-\boldsymbol{y}_1)\cdots\delta^3(\boldsymbol{x}_n-\boldsymbol{y}_n)+\mathrm{sym.}(\boldsymbol{y}_1,\ldots,\boldsymbol{y}_n)\right],\end{aligned} \tag{A.15}$$

where sym.$(\boldsymbol{y}_1,\ldots,\boldsymbol{y}_n)$ represents the symmetric summation of the previous term with respect to $\boldsymbol{y}_1,\ldots,\boldsymbol{y}_n$. If $F[a,a^\dagger]$ is a $c$-number, i.e., the operator which simply multiplies the real number to the operand, then

$$\left\langle F\left[a,a^\dagger\right]\right\rangle_{\mathrm{G}} = \left\langle 0\left|F\left[a,a^\dagger\right]\right|0\right\rangle. \tag{A.16}$$

The quantity (4.53) is, after all, represented as

$$\left\langle \prod_{j=1}^{s} \mathcal{H}_{(m_j)}\left(\boldsymbol{x}_j^{(1)},\ldots,\boldsymbol{x}_j^{(m_j)}\right)\right\rangle_{\mathrm{G}}$$
$$= \left\langle 0 \left| : \prod_{i=1}^{m_1}\left[\tilde{a}\left(\boldsymbol{x}_1^{(i)}\right) + a^\dagger\left(\boldsymbol{x}_1^{(i)}\right)\right] : \cdots : \prod_{i=1}^{m_s}\left[\tilde{a}\left(\boldsymbol{x}_s^{(i)}\right) + a^\dagger\left(\boldsymbol{x}_s^{(i)}\right)\right] : \right| 0 \right\rangle . \quad \text{(A.17)}$$

Expanding the products in *r.h.s*, and using the commutation relations (A.4), (A.5) and the properties (A.11)-(A.13), this quantity is calculated algebraically. But more convenient diagrammatic calculation can be introduced. The "particle" created by a creation operator should be annihilated by an annihilation operator placed left of that creation operator. In a single $\mathcal{H}_{(m_i)}$ represented by Wick product, the creation operators are already placed to the left of annihilation operators, and the "particles" created in some $\mathcal{H}_{(m_i)}$ can not annihilated by the annihilation operator in the same $\mathcal{H}_{(m_i)}$. These observations show the following diagrammatic rules for calculating equation (A.17):

i) Corresponding each $\mathcal{H}_{(m_j)}$, draw $m_j$ points labelled by $\boldsymbol{x}_j^{(1)},\ldots,\boldsymbol{x}_j^{(m_j)}$.

ii) Make $\sum_j m_j/2$ pairs out of those points such that the two points in the same $\mathcal{H}_{(m_j)}$ are not paired. If $\sum_j m_j/2$ is an odd number, the equation (A.17) vanishes.

iii) Associate factor $\psi(\boldsymbol{x}_a^{(p)},\boldsymbol{x}_b^{(q)})$ for each pair, $\boldsymbol{x}_a^{(p)}$ and $\boldsymbol{x}_b^{(q)}$ and make products of these factors.

iv) Sum up those products obtained from all the possible pairings.

For example, the quantities,

$$\left\langle \mathcal{H}_{(1)}(\boldsymbol{x})\mathcal{H}_{(1)}(\boldsymbol{y})\mathcal{H}_{(2)}(\boldsymbol{z},\boldsymbol{w})\right\rangle_{\mathrm{G}}, \quad \text{(A.18)}$$

are evaluated by the diagrams in Figure 12. Applying the above rules we obtain

$$\psi(\boldsymbol{x},\boldsymbol{z})\psi(\boldsymbol{y},\boldsymbol{w}) + \psi(\boldsymbol{x},\boldsymbol{w})\psi(\boldsymbol{y},\boldsymbol{z}). \quad \text{(A.19)}$$

When the generalized Wiener-Hermite functionals of type $\mathcal{H}^{(m)}$ are contained in the *l.h.s.* of equation (A.17), the factor $\psi(\boldsymbol{x}_a^{(p)},\boldsymbol{x}_b^{(q)})$ or $\delta^3(\boldsymbol{x}_a^{(p)}-\boldsymbol{x}_b^{(q)})$ or $\psi^{-1}(\boldsymbol{x}_a^{(p)},\boldsymbol{x}_b^{(q)})$ are associated for each pairs and which factor should be associated can be determined by the fact that the result should be transformed correctly under linear transformation (2.19).

The following orthogonality relation can be derived by the above rules:

$$\left\langle \mathcal{H}_{(n)}(\boldsymbol{x}_1,\ldots,\boldsymbol{x}_n)\mathcal{H}^{(m)}(\boldsymbol{y}_1,\ldots,\boldsymbol{y}_m)\right\rangle_{\mathrm{G}}$$
$$= \delta_{mn}\left[\delta^3(\boldsymbol{x}_1-\boldsymbol{y}_1)\cdots\delta^3(\boldsymbol{x}_m-\boldsymbol{y}_m) + \mathrm{sym.}(\boldsymbol{y}_1,\ldots,\boldsymbol{y}_m)\right]. \quad \text{(A.20)}$$

## A2 The method using generating functional

The second method using generating functionals is to derive rules i) $\sim$ iv) using generating functional of generalized Wiener-Hermite functionals. This method is similar to the derivation of Feynman rules in path integral formalism in quantum field theory.

The generating functional of generalized Wiener-Hermite functionals is defined by

$$\mathcal{G}[\alpha, J] = \sum_{m=0}^{\infty} \frac{1}{m!} \int d^3x_1 \cdots d^3x_m J(\boldsymbol{x}_1) \cdots J(\boldsymbol{x}_m) \mathcal{H}^{(m)}(\boldsymbol{x}_1, \ldots, \boldsymbol{x}_m), \tag{A.21}$$

so that

$$\mathcal{H}^{(m)}(\boldsymbol{x}_1, \ldots, \boldsymbol{x}_m) = \left. \frac{\delta^m \mathcal{G}[\alpha, J]}{\delta J(\boldsymbol{x}_1) \cdots \delta J(\boldsymbol{x}_m)} \right|_{J=0}. \tag{A.22}$$

We consider the following functional:

$$F[J-\alpha] = \exp\left\{ -\frac{1}{2} \int d^3x d^3y [J(\boldsymbol{x}) - \alpha(\boldsymbol{x})] \psi^{-1}(\boldsymbol{x}, \boldsymbol{y}) [J(\boldsymbol{y}) - \alpha(\boldsymbol{y})] \right\}, \tag{A.23}$$

and expand this functional with respect to $J$:

$$F[J-\alpha] = \sum_{m=0}^{\infty} \frac{1}{m!} \int d^3x_1 \cdots d^3x_m J(\boldsymbol{x}_1) \cdots J(\boldsymbol{x}_m) \left. \frac{\delta^m F[J-\alpha]}{\delta J(\boldsymbol{x}_1) \cdots \delta J(\boldsymbol{x}_m)} \right|_{J=0}. \tag{A.24}$$

Because

$$\begin{aligned} \left. \frac{\delta^m F[J-\alpha]}{\delta J(\boldsymbol{x}_1) \cdots \delta J(\boldsymbol{x}_m)} \right|_{J=0} &= (-1)^m \left. \frac{\delta^m F[J-\alpha]}{\delta \alpha(\boldsymbol{x}_1) \cdots \delta \alpha(\boldsymbol{x}_m)} \right|_{J=0} \\ &= (-1)^m \left. \frac{\delta^m F[\alpha]}{\delta \alpha(\boldsymbol{x}_1) \cdots \delta \alpha(\boldsymbol{x}_m)} \right|_{J=0} = F[\alpha] \mathcal{H}^{(m)}(\boldsymbol{x}_1, \ldots, \boldsymbol{x}_m), \end{aligned} \tag{A.25}$$

we see

$$\begin{aligned} \mathcal{G}[\alpha, J] &= \frac{F[J-\alpha]}{F[\alpha]} \\ &= \exp\left\{ -\frac{1}{2} \int d^3x d^3y J(\boldsymbol{x}) \psi^{-1}(\boldsymbol{x}, \boldsymbol{y}) J(\boldsymbol{y}) + \int d^3x d^3y \alpha(\boldsymbol{x}) \psi^{-1}(\boldsymbol{x}, \boldsymbol{y}) J(\boldsymbol{y}) \right\} \end{aligned} \tag{A.26}$$

Using this representation of generating functional, we can derive the following equation:

$$\langle \mathcal{G}[\alpha, J_1] \cdots \mathcal{G}[\alpha, J_s] \rangle_{\mathrm{G}} = \exp\left[ \sum_{i<j}^{s} \int d^3x d^3y J_i(\boldsymbol{x}) \psi^{-1}(\boldsymbol{x}, \boldsymbol{y}) J_j(\boldsymbol{y}) \right]. \tag{A.27}$$

The rules i) $\sim$ iv) are derived by operating

$$\frac{\delta^{m_1}}{\delta J_1\left(\boldsymbol{x}_1^{(1)}\right) \cdots \delta J_1\left(\boldsymbol{x}_1^{(m_1)}\right)} \cdots \frac{\delta^{m_s}}{\delta J_s\left(\boldsymbol{x}_s^{(1)}\right) \cdots \delta J_s\left(\boldsymbol{x}_s^{(m_s)}\right)}, \tag{A.28}$$

to the both side of equation (A.27), and putting $J_1 = J_2 = \cdots = J_s = 0$.

## APPENDIX B: THE CALCULATION OF THE STATISTICAL FACTOR

In the expression (3.5) the points associated with $\mathcal{H}^{(n_1+\cdots+n_m)}, \mathcal{H}_{(m_i)}$ are classified into $m+N$ groups with $n_1, \ldots, n_m; m_1, \ldots, m_N$ members respectively. These points in each group are symmetric about permutations. We distinguish these $m+N$ groups by labels like $\alpha, \beta$, and so on. In evaluating the Gaussian average of the product of generalized Wiener-Hermite functionals using diagrammatic method described in Appendix A, we denote the number of lines connecting the members of two groups $\alpha$ and $\beta$ as a symmetric matrix $M_{\alpha\beta}$ with vanishing diagonal elements. Obviously, graphs which have the same value of the matrix $M_{\alpha\beta}$ give the same contribution to $P_N$. Thus we need to calculate the degree of multiplicity of the graphs which have the same value of $M_{\alpha\beta}$.

We designate the number of points in a group $\alpha$ as $M_\alpha$. The number of ways to divide the $M_\alpha$ points into $M_{\alpha,1}, \ldots, M_{\alpha,m+N}$ parts for all groups $\alpha$'s is

$$\prod_\alpha \left( \frac{M_\alpha!}{M_{\alpha,1}! \cdots M_{\alpha,m+N}!} \right) = \frac{\prod_\alpha M_\alpha!}{\left( \prod_{\alpha<\beta} M_{\alpha\beta}! \right)^2}. \tag{B.1}$$

The number of ways to connect these points is given by $\prod_{\alpha<\beta} M_{\alpha\beta}!$ and thus the multiplicity of the graphs is

$$\frac{n_1! \cdots n_m! m_1! \cdots m_N!}{\prod_{\text{ends}} a_{\text{ends}}}. \tag{B.2}$$

The numerator cancels out in equation (3.5) and the statistical factor (3.6) follows. It is obvious that the $m$ vertices are distinguished in the above calculations and the graphs like Figure 13 are counted as different graphs.

## APPENDIX C: EVALUATION OF $R_m^{(\mathrm{pk})}(\nu)$

This Appendix is devoted to the derivation of equation (4.15). The following identity is convenient for our purpose:

$$\frac{x^2}{2} + \frac{(y-\gamma x)^2}{2(1-\gamma^2)} = \frac{y^2}{2} + \frac{(x-\gamma y)^2}{2(1-\gamma^2)}. \tag{C.1}$$

From this identity, equation (4.8) reduces to

$$\mathcal{N}_{\mathrm{pk}}^{(\mathrm{G})}(\nu) = \frac{1}{(2\pi)^2 R_*^3} \int_0^\infty dx f(x) e^{-x^2/2} \frac{\exp\left[ -\frac{(\nu-\gamma x)^2}{2(1-\gamma^2)} \right]}{[2\pi(1-\gamma^2)]^{1/2}}, \tag{C.2}$$

and the integral in equation (4.7) can easily be calculated:

$$\mathrm{n}_{\mathrm{pk}}^{(\mathrm{G})}(\nu) = \frac{1}{8\pi^2 R_*^3}\int_0^\infty dx f(x) e^{-x^2/2}\mathrm{erfc}\left(\frac{\nu-\gamma x}{\sqrt{2(1-\gamma^2)}}\right). \tag{C.3}$$

The latter expression (C.3) was previously derived by Mann, Heavens & Peecock (1993). Differentiating equation (C.2) and using the identity (C.1) again, we obtain

$$\left(-\frac{d}{d\nu}\right)^{m-1} \mathcal{N}_{\mathrm{pk}}^{(\mathrm{G})}(\nu) = \frac{e^{-\nu^2/2}}{(2\pi)^{5/2}R_*^3(1-\gamma^2)^{m/2}}\int_0^\infty dx f(x) H_{m-1}\left(\frac{\nu-\gamma x}{\sqrt{1-\gamma^2}}\right)\exp\left[-\frac{(x-\gamma\nu)^2}{2(1-\gamma^2)}\right], \tag{C.4}$$

for $m \geq 1$. Thus,

$$R_m^{(\mathrm{pk})}(\nu) = \begin{cases} \dfrac{1}{8\pi^2 R_*^3}\displaystyle\int_0^\infty dx f(x) e^{-x^2/2}\mathrm{erfc}\left(\frac{\nu-\gamma x}{\sqrt{2(1-\gamma^2)}}\right) & (m=0) \\ \dfrac{e^{-\nu^2/2}}{(2\pi)^{5/2}R_*^3(1-\gamma^2)^{m/2}}\displaystyle\int_0^\infty dx f(x) H_{m-1}\left(\frac{\nu-\gamma x}{\sqrt{1-\gamma}}\right)\exp\left[-\frac{(x-\gamma\nu)^2}{2(1-\gamma^2)}\right] & (m \geq 1) \end{cases} \tag{C.5}$$

is proved.

## APPENDIX D: PROOF OF $\langle\tilde{\alpha}\rangle = \mathcal{N}_{\mathrm{pk}}^{(\mathrm{G})}(\nu)/\mathrm{n}_{\mathrm{pk}}^{(\mathrm{G})}(\nu)$

In this Appendix, we calculate $\langle\tilde{\alpha}\rangle$ defined by equation (4.19). Using the identity (C.1), the equation (4.19) reduces to

$$\langle\tilde{\alpha}\rangle = \frac{1}{(2\pi)^2 R_*^3 \mathrm{n}_{\mathrm{pk}}^{(\mathrm{G})}(\nu)}\int_0^\infty dx f(x) e^{-x^2/2}\int_\nu^\infty d\alpha \frac{\exp\left[-\dfrac{(\alpha-\gamma x)^2}{2(1-\gamma^2)}\right]}{[2\pi(1-\gamma^2)]^{1/2}}\frac{\alpha-\gamma x}{1-\gamma^2}. \tag{D.1}$$

The integration with respect to $\alpha$ is straightforward by a transformation $\beta = \alpha - \gamma x$. Comparing the result with equation (C.2), this quantity has the simple form,

$$\langle\tilde{\alpha}\rangle = \frac{\mathcal{N}_{\mathrm{pk}}^{(\mathrm{G})}(\nu)}{\mathrm{n}_{\mathrm{pk}}^{(\mathrm{G})}(\nu)}. \tag{D.2}$$

## APPENDIX E: DERIVATION OF COEFFICIENTS FOR DENSITY EXTREMA

In this Appendix, we derive the coefficients (4.26). The quantity we should calculate is

$$
\begin{aligned}
&R(k,\{l_i\},\{p_{ij}\}) = \\
&\qquad \left\langle \left(\frac{\partial}{\partial\alpha}\right)^k \prod_{i=1}^{3}\left(\frac{\partial}{\partial\beta_i}\right)^{l_i} \prod_{i\le j}^{3}\left(\frac{\partial}{\partial\omega_{ij}}\right)^{p_{ij}} \theta(\alpha-\nu)\delta^3(\beta_i)(-1)^3\det(\omega_{ij})\right\rangle_{\mathrm{G}} \qquad \text{(E.1)}\\
&\qquad = -\frac{1}{(2\pi)^5\sqrt{\det M}}\int_{-\infty}^{\infty} d\alpha \prod_{i=1}^{3} d\beta_i \prod_{i\le j}^{3} d\omega_{ij} \exp\left[-\frac{1}{2}\sum_{\mu,\nu} A_\mu (M^{-1})_{\mu\nu} A_\nu\right]\\
&\qquad\qquad \times \frac{\partial^k\theta(\alpha-\nu)}{\partial\alpha^k}\prod_{i=1}^{3}\frac{\partial^{l_i}\delta(\beta_i)}{\partial\beta_i^{l_i}}\prod_{i\le j}^{3}\left(\frac{\partial}{\partial\omega_{ij}}\right)^{p_{ij}}\det\omega, \qquad \text{(E.2)}
\end{aligned}
$$

where ten dimensional vector $A_\mu$ is

$$
(A_\mu) = (\alpha, \beta_1, \beta_2, \beta_3, \omega_{11}, \omega_{22}, \omega_{33}, \omega_{23}, \omega_{13}, \omega_{12}). \qquad \text{(E.3)}
$$

The correlation matrix $M_{\mu\nu} = \langle A_\mu A_\nu \rangle$ is explicitly

$$
M = \left[\begin{array}{c|c|c|c}
1 & \boldsymbol{O}_V^T & -\frac{\sigma_1^2}{3\sigma_0^2}\boldsymbol{I}_V^T & \boldsymbol{O}_V^T \\ \hline
\boldsymbol{O}_V & \frac{\sigma_1^2}{3\sigma_0^2}\boldsymbol{I} & \boldsymbol{O} & \boldsymbol{O} \\ \hline
-\frac{\sigma_1^2}{3\sigma_0^2}\boldsymbol{I}_V & \boldsymbol{O} & \frac{\sigma_2^2}{15\sigma_0^2}\boldsymbol{B} & \boldsymbol{O} \\ \hline
\boldsymbol{O}_V & \boldsymbol{O} & \boldsymbol{O} & \frac{\sigma_2^2}{15\sigma_0^2}\boldsymbol{I}
\end{array}\right], \qquad \text{(E.4)}
$$

where

$$
\boldsymbol{O}_V = \begin{bmatrix} 0\\0\\0 \end{bmatrix}, \boldsymbol{I}_V = \begin{bmatrix} 1\\1\\1 \end{bmatrix}, \boldsymbol{O} = \begin{bmatrix} 0&0&0\\0&0&0\\0&0&0 \end{bmatrix}, \boldsymbol{I} = \begin{bmatrix} 1&0&0\\0&1&0\\0&0&1 \end{bmatrix}, \boldsymbol{B} = \begin{bmatrix} 3&1&1\\1&3&1\\1&1&3 \end{bmatrix}. \qquad \text{(E.5)}
$$

This is the consequence of spatial homogeneity and isotropy (see Bardeen et al. 1986). We introduce new variables $\tilde{\omega}_{ij}$ as

$$
\tilde{\omega}_{ij} = \omega_{ij} + \delta_{ij}\frac{\sigma_1^2}{3\sigma_0^2}\alpha, \qquad \text{(E.6)}
$$

then the non-vanishing correlations among variables $\alpha$, $\beta_i$, $\tilde{\omega}_{ij}$ are

$$
\langle\alpha^2\rangle = 1, \langle\beta_1^2\rangle = \langle\beta_2^2\rangle = \langle\beta_3^2\rangle = \frac{\sigma_1^2}{3\sigma_0^2},
$$

$$
\begin{aligned}
\langle\tilde{\omega}_{11}^2\rangle = \langle\tilde{\omega}_{22}^2\rangle = \langle\tilde{\omega}_{33}^2\rangle &= \frac{\sigma_2^2}{5\sigma_0^2}\left(1-\frac{5}{9}\gamma^2\right), \\
\langle\tilde{\omega}_{11}\tilde{\omega}_{22}\rangle = \langle\tilde{\omega}_{22}\tilde{\omega}_{33}\rangle = \langle\tilde{\omega}_{11}\tilde{\omega}_{33}\rangle &= \frac{\sigma_2^2}{15\sigma_0^2}\left(1-\frac{5}{3}\gamma^2\right), \\
\langle\tilde{\omega}_{12}^2\rangle = \langle\tilde{\omega}_{23}^2\rangle = \langle\tilde{\omega}_{13}^2\rangle &= \frac{\sigma_2^2}{15\sigma_0^2},
\end{aligned}
\tag{E.7}
$$

and all other correlations vanish. The coefficients (E.1) is represented by new variables, $\alpha$, $\beta_i$, $\omega_{ij}$ as

$$
R(k,\{l_i\},\{p_{ij}\}) = -\left\langle \prod_{i=1}^{3} \frac{\partial^{l_i}\delta(\beta_i)}{\partial\beta_i^{l_i}} \right\rangle_{\mathrm{G}} \left\langle \frac{\partial^k\theta(\alpha-\nu)}{\partial\alpha^k} \prod_{i\le j}^{3} \left(\frac{\partial}{\partial\tilde{\omega}_{ij}}\right)^{p_{ij}} W(\alpha,\tilde{\omega}) \right\rangle_{\mathrm{G}}, \tag{E.8}
$$

where

$$
\begin{aligned}
W(\alpha,\tilde{\omega}) = &-\left(\frac{\sigma_1^2}{3\sigma_0^2}\right)^3\alpha^3 + \left(\frac{\sigma_1^2}{3\sigma_0^2}\right)^2 (\tilde{\omega}_{11}+\tilde{\omega}_{22}+\tilde{\omega}_{33})\,\alpha^2 \\
&+\frac{\sigma_1^2}{3\sigma_0^2}\left(\tilde{\omega}_{23}^2+\tilde{\omega}_{13}^2+\tilde{\omega}_{12}^2-\tilde{\omega}_{22}\tilde{\omega}_{33}-\tilde{\omega}_{11}\tilde{\omega}_{33}-\tilde{\omega}_{11}\tilde{\omega}_{22}\right)\alpha \\
&+\tilde{\omega}_{11}\tilde{\omega}_{22}\tilde{\omega}_{33}+2\tilde{\omega}_{23}\tilde{\omega}_{13}\tilde{\omega}_{12}-\tilde{\omega}_{11}\tilde{\omega}_{23}^2-\tilde{\omega}_{22}\tilde{\omega}_{13}^2-\tilde{\omega}_{33}\tilde{\omega}_{12}^2.
\end{aligned}
\tag{E.9}
$$

We see, from equation (E.9), that the equation (E.8) survives only when integers $p_{ij}$ take the values listed in Table 1. Using $J_i(\{p_{ij}\})$ defined by Table 1, the multivariate Gaussian integral by $\tilde{\omega}$ of the second factor of the equation (E.8) reduces to

$$
\begin{aligned}
-\left(\frac{\sigma_1^2}{3\sigma_0^2}\right)^{3-\sum_{i\le j}p_{ij}} \Bigg\langle \frac{\partial^k\theta(\alpha-\nu)}{\partial\alpha^k} &\left[H_3(\alpha)J_0(\{p_{ij}\}) - H_2(\alpha)J_1(\{p_{ij}\}) + \right. \\
&\left. + H_1(\alpha)J_2(\{p_{ij}\}) - J_3(\{p_{ij}\})\right] \Bigg\rangle_{\mathrm{G}}.
\end{aligned}
\tag{E.10}
$$

Defining

$$
H_{-1}(\nu) = \sqrt{\frac{\pi}{2}}e^{\nu^2/2}\mathrm{erfc}\left(\frac{\nu}{\sqrt{2}}\right), \tag{E.11}
$$

we find

$$
\left\langle \frac{\partial^k\theta(\alpha-\nu)}{\partial\alpha^k} H_n(\alpha) \right\rangle_{\mathrm{G}} = \frac{1}{\sqrt{2\pi}} H_{k+n-1}(\nu)e^{-\nu^2/2}, \tag{E.12}
$$

for $n = 0, 1, 2, \ldots$. The first factor of equation (E.8) is

$$
\frac{1}{(2\pi)^{3/2}}\left(\frac{\sigma_1}{\sqrt{3}\sigma_0}\right)^{-\sum_i l_i - 3} H_{l_1}(0)H_{l_2}(0)H_{l_3}(0), \tag{E.13}
$$

where

$$H_l(0) = \left\{ \begin{array}{ll} 0 & (l\text{: odd}), \\ (-1)^{l/2}(l-1)!! & (l\text{: even}). \end{array} \right. . \tag{E.14}$$

Therefore, we finally get the explicit form of the kernels:

$$\begin{aligned} R(k, \{l_i\}, \{p_{ij}\}) = & \frac{1}{(2\pi)^2} \left( \frac{\sigma_1}{\sqrt{3}\sigma_0} \right)^{3 - \sum_i l_i - 2\sum_{i \le j} p_{ij}} H_{l_1}(0) H_{l_2}(0) H_{l_3}(0) \\ & \times e^{-\nu^2/2} \left[ H_{k+2}(\nu) J_0(\{p_{ij}\}) - H_{k+1}(\nu) J_1(\{p_{ij}\}) \right. \\ & \left. + H_k(\nu) J_2(\{p_{ij}\}) - H_{k-1}(\nu) J_3(\{p_{ij}\}) \right] . \end{aligned} \tag{E.15}$$

| $p_{11}$ | $p_{22}$ | $p_{33}$ | $p_{23}$ | $p_{13}$ | $p_{12}$ | $J_0$ | $J_1$ | $J_2$ | $J_3$ |
|---|---|---|---|---|---|---|---|---|---|
| 0 | 0 | 0 | 0 | 0 | 0 | 1 | 0 | 0 | 0 |
| 1 | 0 | 0 | 0 | 0 | 0 | 0 | 1 | 0 | 0 |
| 0 | 1 | 0 | 0 | 0 | 0 | 0 | 1 | 0 | 0 |
| 0 | 0 | 1 | 0 | 0 | 0 | 0 | 1 | 0 | 0 |
| 1 | 1 | 0 | 0 | 0 | 0 | 0 | 0 | 1 | 0 |
| 1 | 0 | 1 | 0 | 0 | 0 | 0 | 0 | 1 | 0 |
| 0 | 1 | 1 | 0 | 0 | 0 | 0 | 0 | 1 | 0 |
| 0 | 0 | 0 | 2 | 0 | 0 | 0 | 0 | $-2$ | 0 |
| 0 | 0 | 0 | 0 | 2 | 0 | 0 | 0 | $-2$ | 0 |
| 0 | 0 | 0 | 0 | 0 | 2 | 0 | 0 | $-2$ | 0 |
| 1 | 1 | 1 | 0 | 0 | 0 | 0 | 0 | 0 | 1 |
| 1 | 0 | 0 | 2 | 0 | 0 | 0 | 0 | 0 | $-2$ |
| 0 | 1 | 0 | 0 | 2 | 0 | 0 | 0 | 0 | $-2$ |
| 0 | 0 | 1 | 0 | 0 | 2 | 0 | 0 | 0 | $-2$ |
| 0 | 0 | 0 | 1 | 1 | 1 | 0 | 0 | 0 | 2 |

Table 1: The definition of $J_i$. For other values of $p_{ij}$ not listed in this table, $J_i = 0$.

## FIGURE CAPTIONS

Figure 1 : Diagrammatic rules for the nonlocal bias. The open circles and filled circles corresponds to external points and vertices, respectively.

Figure 2 : Diagrammatic rules for the local bias.

Figure 3 : Diagrammatic rules for the semi-local bias.

Figure 4 : Diagrammatic rules in Fourier space.

Figure 5 : Diagrams for the Edgeworth expansion.

Figure 6 : Various approximations for the two-point correlation function of peaks in CDM model ($h = 0.5$) for thresholds $\nu = 1.5$, $\nu = 2.0$, $\nu = 2.5$, and $\nu = 3.0$. The smoothing length is $0.2h^{-1}$Mpc and the Gaussian window function is adopted. *Solid lines*: our method. *Dotted lines*: Jensen and Szalay (1986). *Short Dash lines*: peak-background splitting. *Long dash lines*: Bardeen et al. (1986). *Dot-short dash lines*: Lumsden et al. (1989). *Dot-long dash lines*: original correlation function of CDM model for reference. Normalization is not relevant for our analysis.

Figure 7 : Various approximations for the three-point correlation function of peaks in CDM model. The configurations of three-points is equilateral and the length of edges of the equilateral triangle is the horizontal axis. *Solid lines*: our method. *Dotted lines*: Jensen and Szalay (1986). *Short Dash lines*: peak-background splitting. *Dot-short dash lines*: Jensen and Szalay with effective thresholds.

Figure 8 : The normalized three-point correlation function of equilateral configuration for the same approximations as in Figure 7.

Figure 9 : Diagrams for $n_{\rm ex.}$ and $\xi_{{\rm ex},\nu}$.

Figure 10 : Diagrams for hierarchical underlying fluctuation.

Figure 11 : Definitions for tree graphs $\alpha$, $\beta$, $\gamma$.

Figure 12 : An example of diagrams for expectation values of products of generalized Wiener-Hermite functionals.

Figure 13 : These two graphs are counted separately for the statistical factor described in the text.

FIG. 1

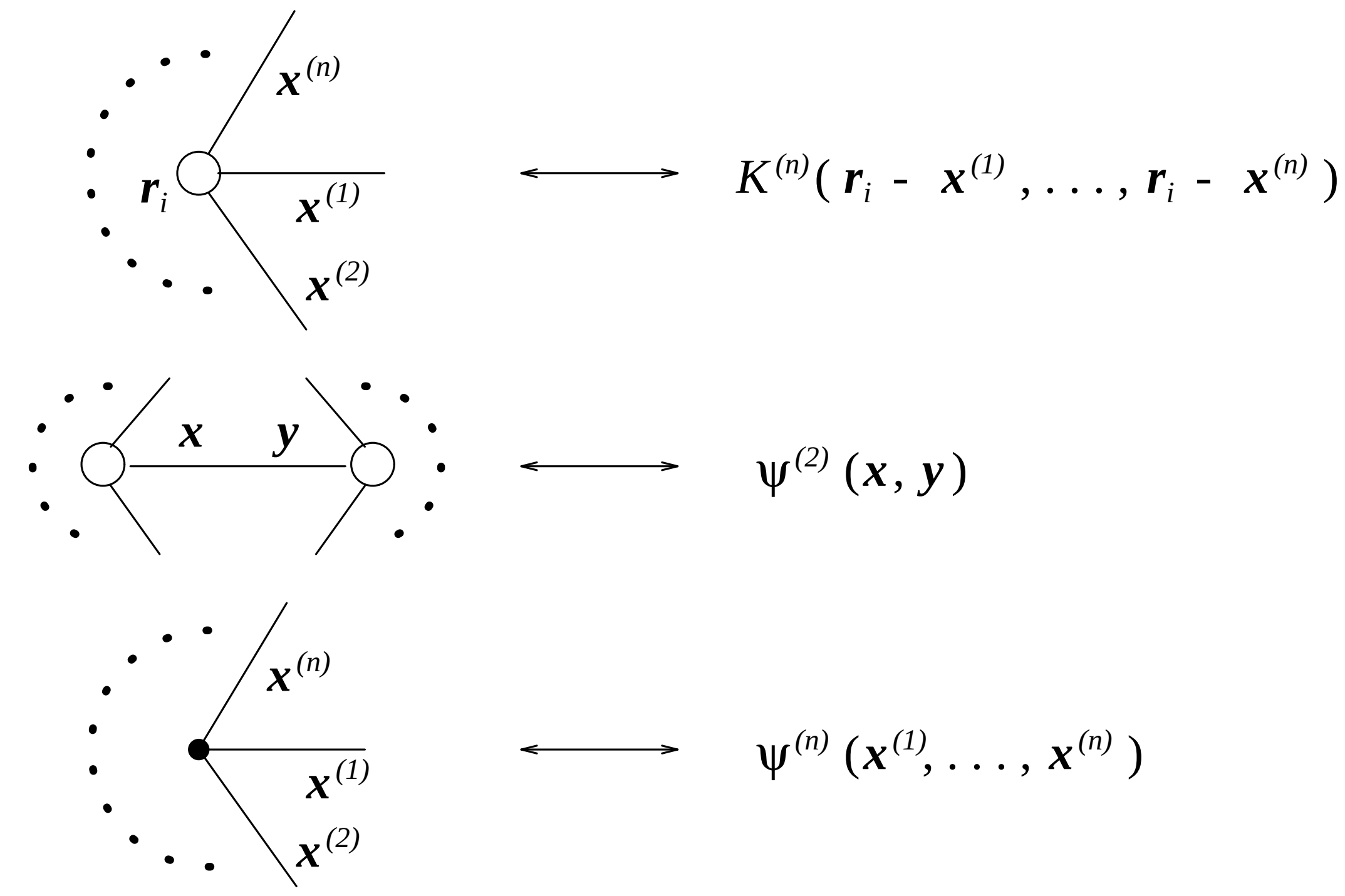
x^(n)
r_i
x^(1)
x^(2)
K^(n)( r_i - x^(1) , . . . , r_i - x^(n) )
x
y
ψ^(2) (x, y)
x^(n)
x^(1)
x^(2)
ψ^(n) (x^(1), . . . , x^(n) )

FIG. 2

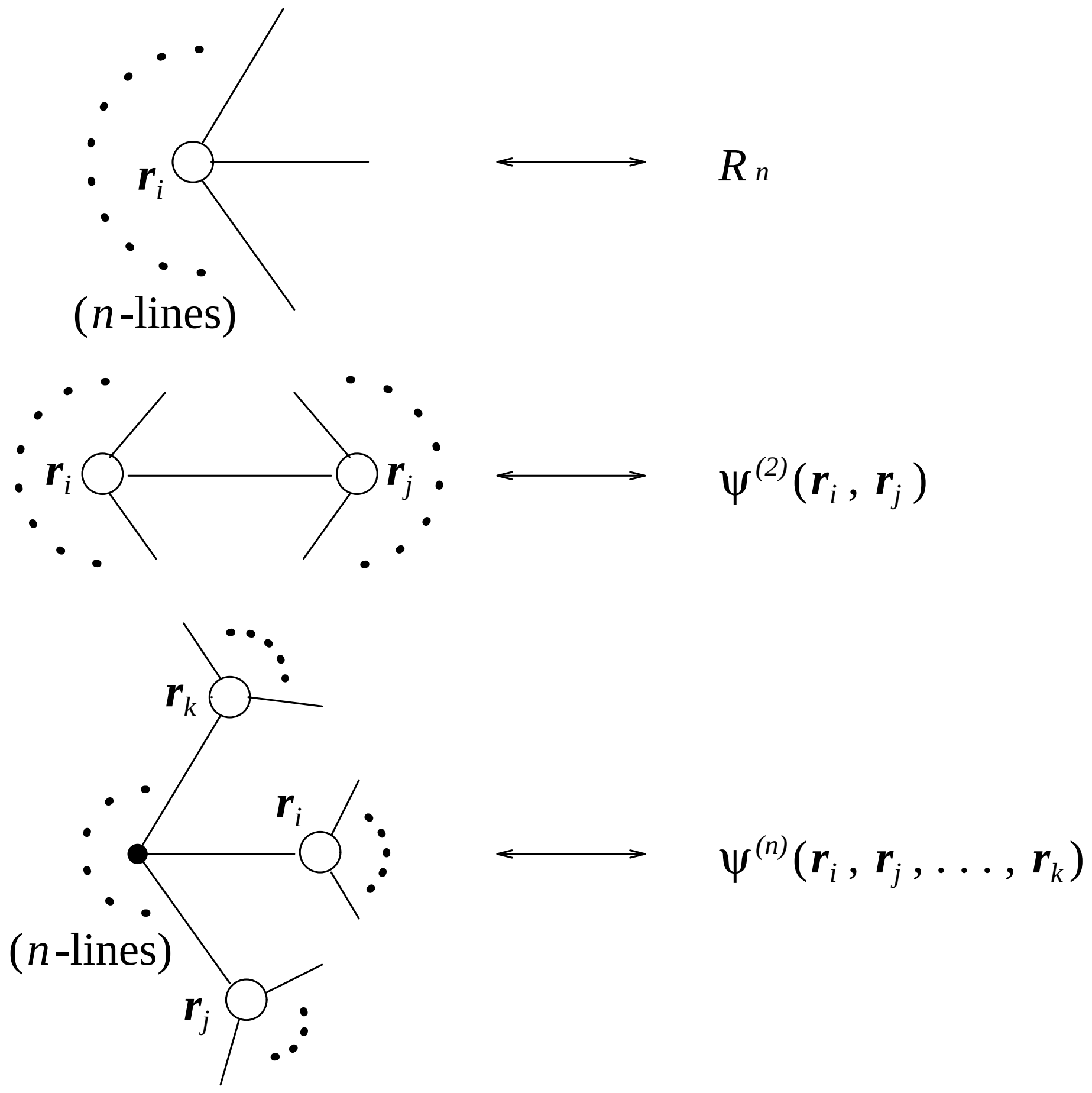
r_i
(n-lines)
R_n
r_i
r_j
ψ^(2)(r_i , r_j )
r_k
r_i
(n-lines)
r_j
ψ^(n)(r_i , r_j , . . . , r_k)

FIG. 3

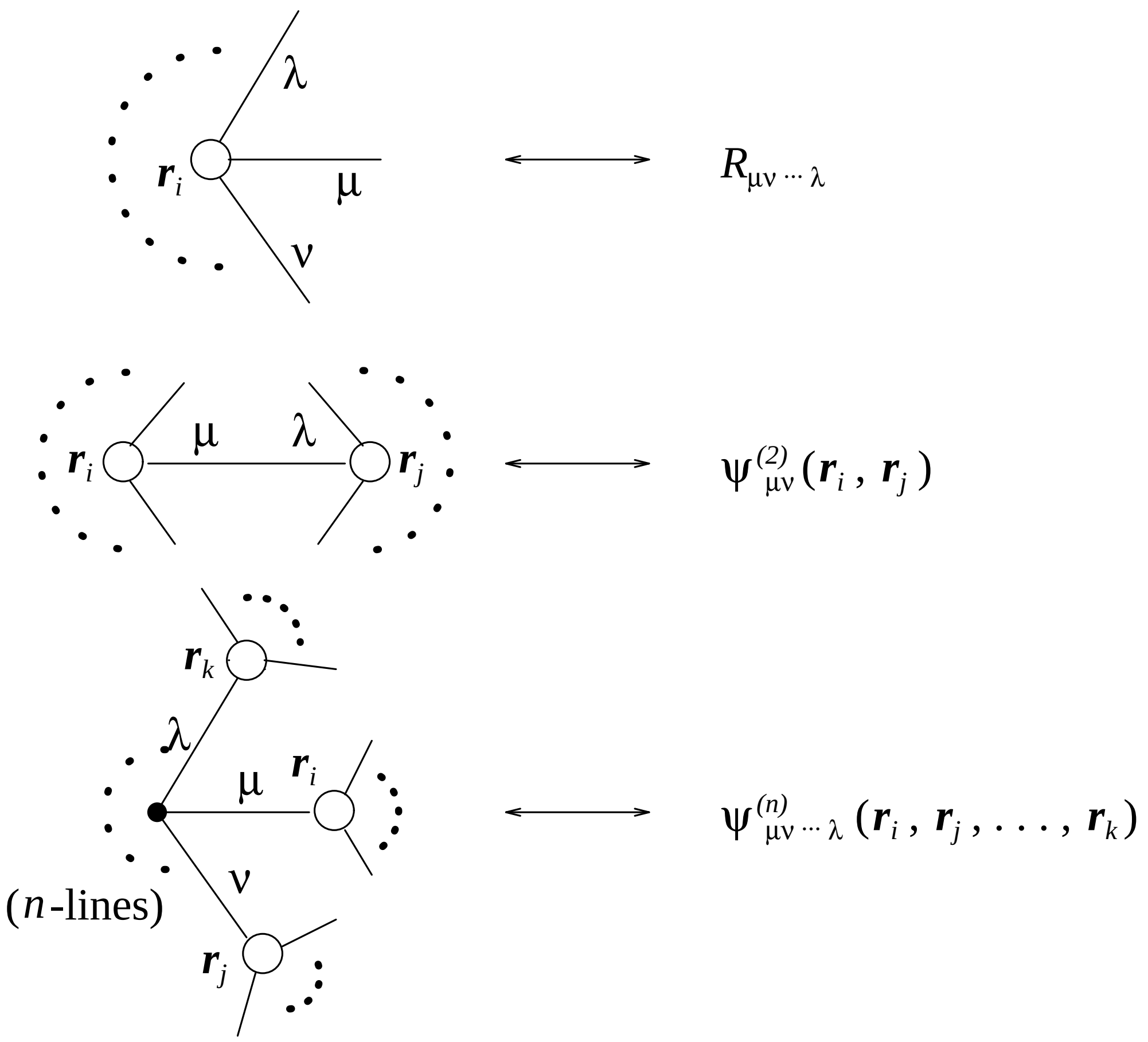
λ
$\boldsymbol{r}_i$
μ
ν
$R_{\mu\nu\cdots\lambda}$
μ
λ
$\boldsymbol{r}_i$
$\boldsymbol{r}_j$
$\psi^{(2)}_{\mu\nu}(\boldsymbol{r}_i, \boldsymbol{r}_j)$
$\boldsymbol{r}_k$
λ
μ
$\boldsymbol{r}_i$
ν
($n$-lines)
$\boldsymbol{r}_j$
$\psi^{(n)}_{\mu\nu\cdots\lambda}(\boldsymbol{r}_i, \boldsymbol{r}_j, \ldots, \boldsymbol{r}_k)$

FIG. 4

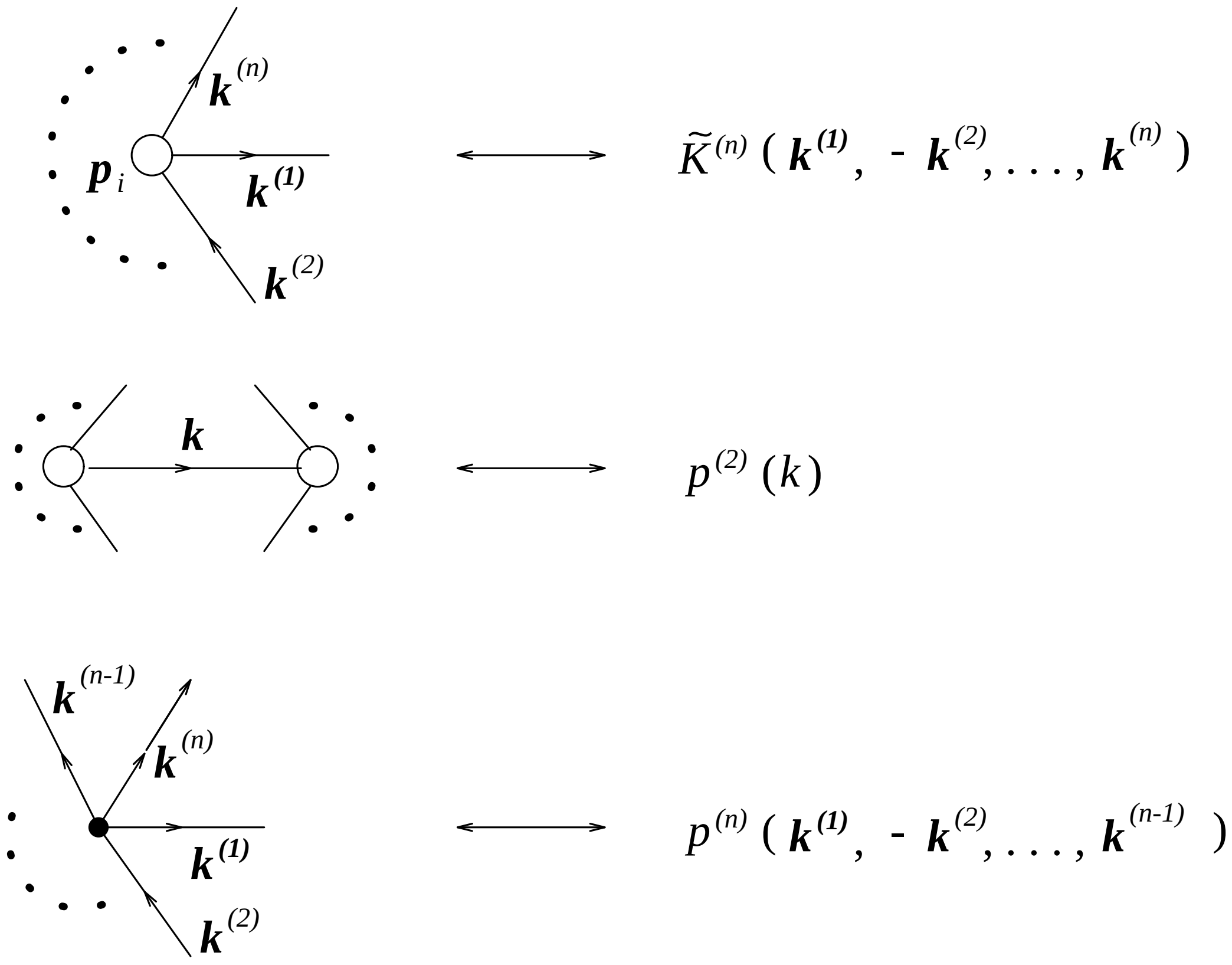
$\boldsymbol{k}^{(n)}$
$\boldsymbol{p}_i$
$\boldsymbol{k}^{(1)}$
$\boldsymbol{k}^{(2)}$
$\widetilde{K}^{(n)}(\boldsymbol{k}^{(1)}, -\boldsymbol{k}^{(2)}, \ldots, \boldsymbol{k}^{(n)})$
$\boldsymbol{k}$
$p^{(2)}(k)$
$\boldsymbol{k}^{(n-1)}$
$\boldsymbol{k}^{(n)}$
$\boldsymbol{k}^{(1)}$
$\boldsymbol{k}^{(2)}$
$p^{(n)}(\boldsymbol{k}^{(1)}, -\boldsymbol{k}^{(2)}, \ldots, \boldsymbol{k}^{(n-1)})$

FIG. 5

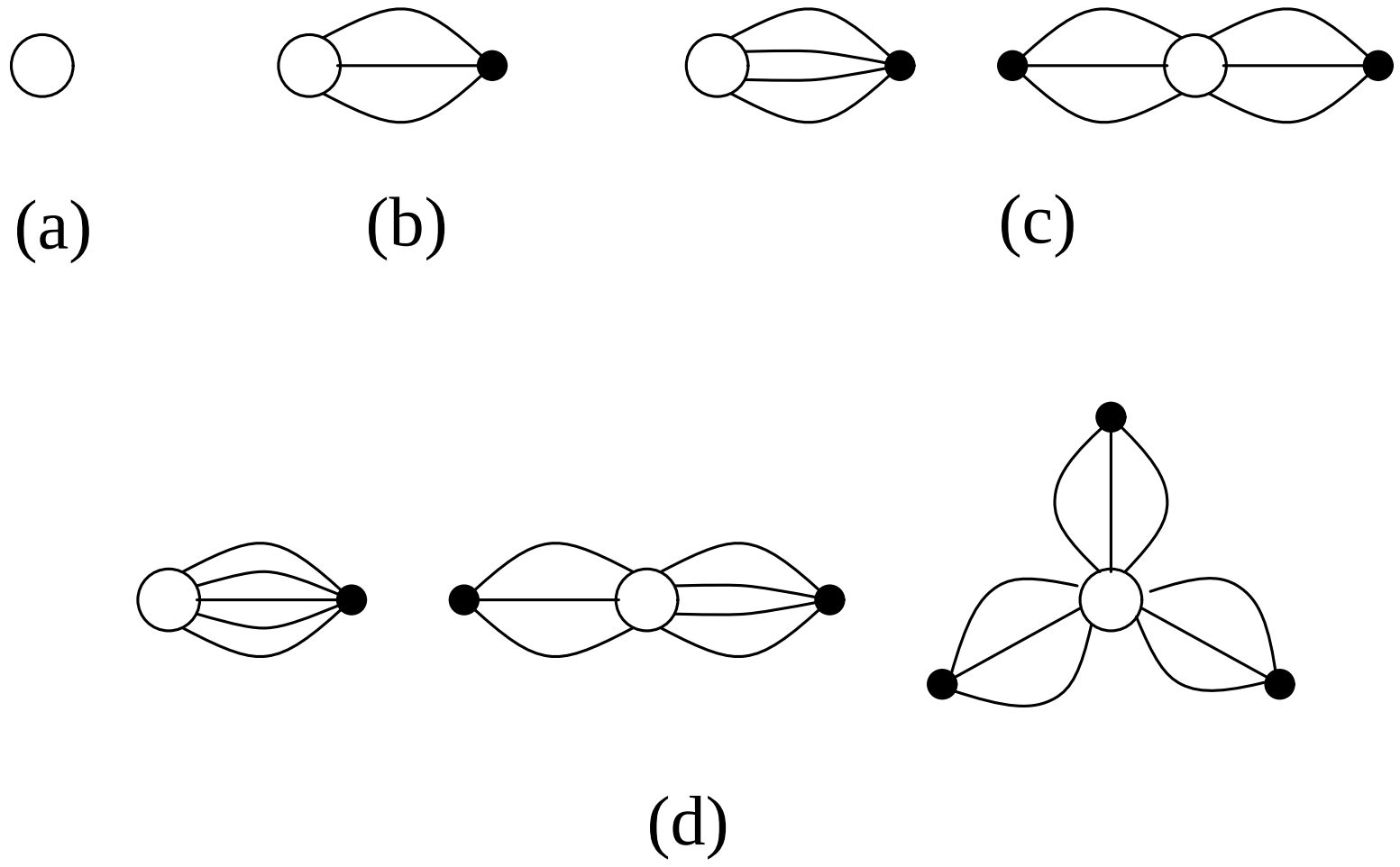

FIG. 6

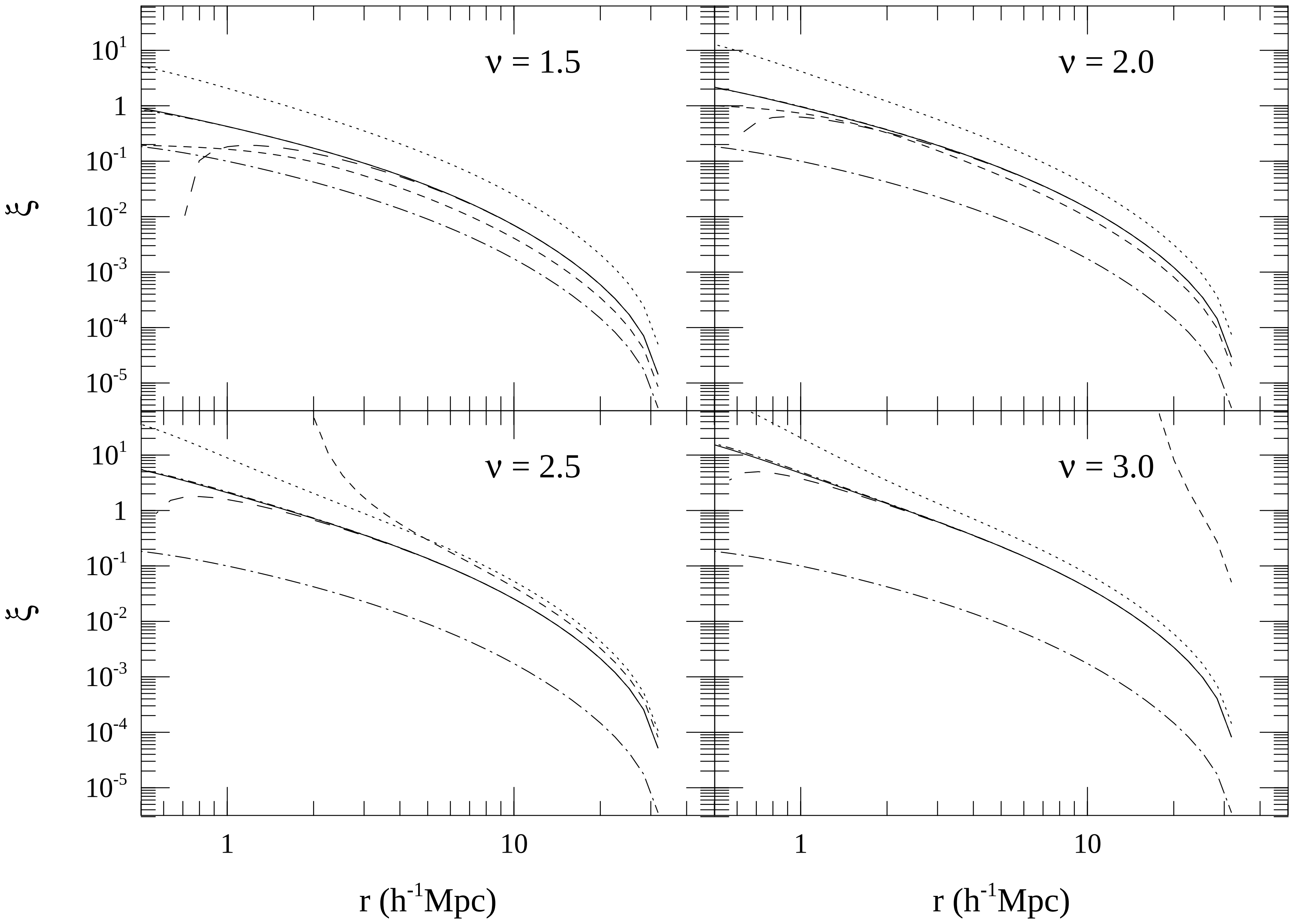

FIG. 7

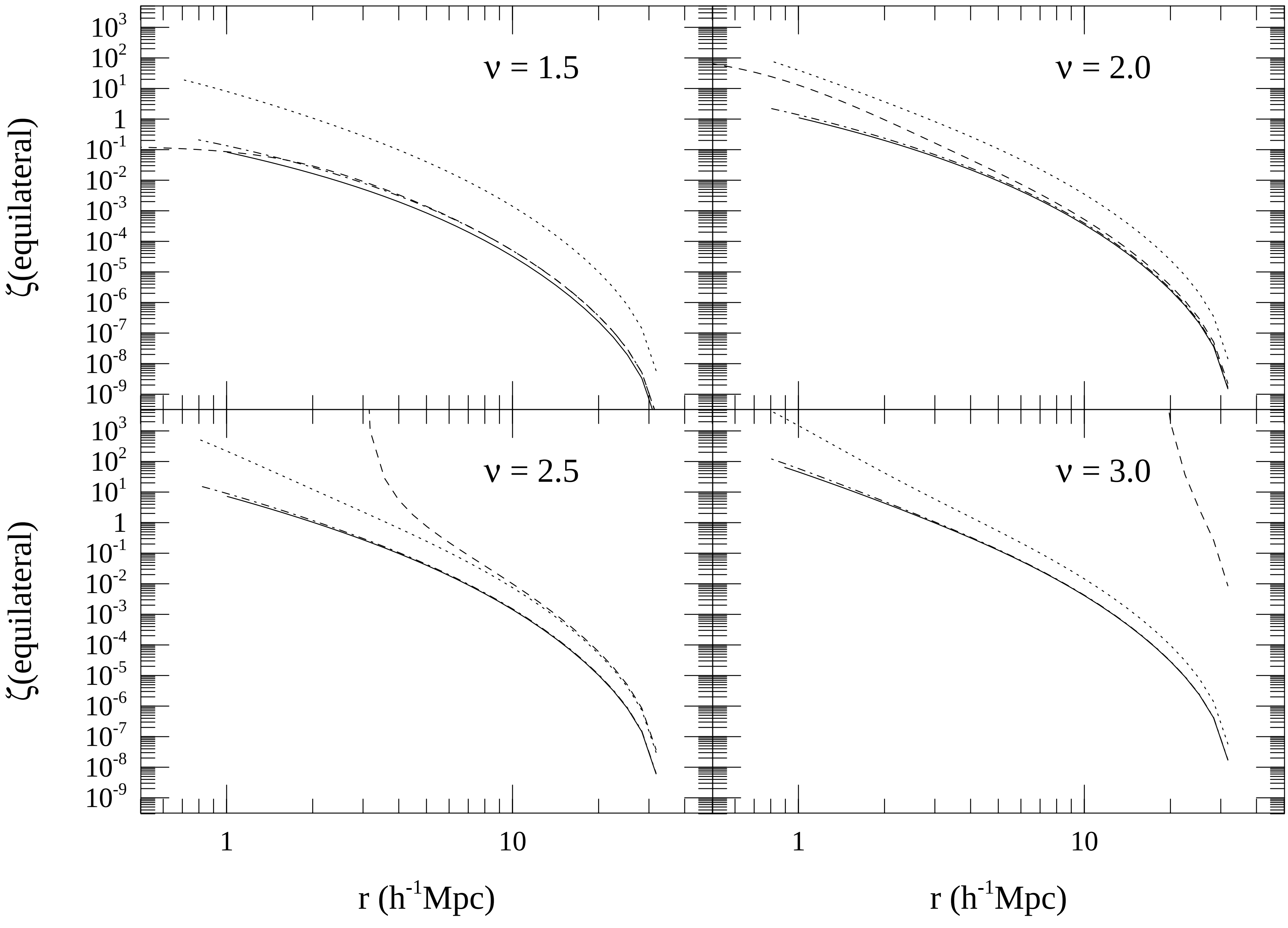

FIG. 8

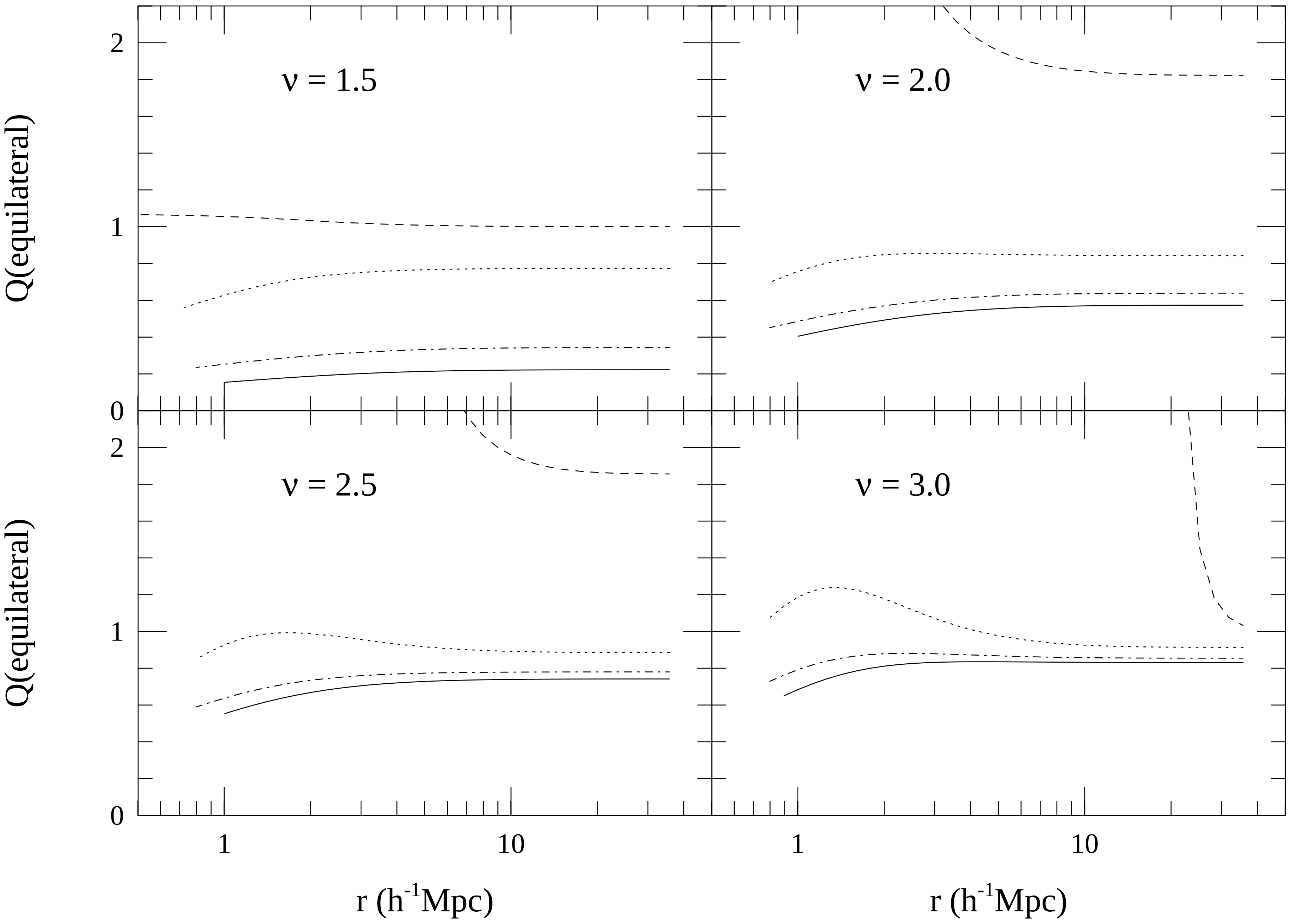

FIG. 9

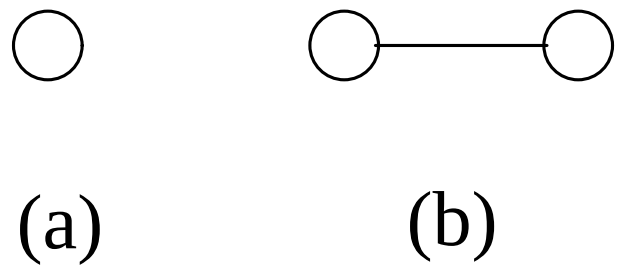

FIG. 10

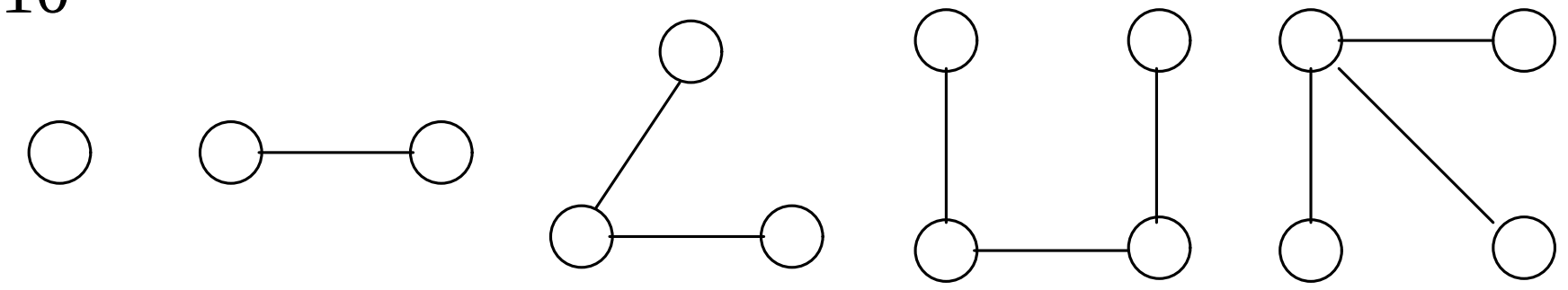

FIG. 11

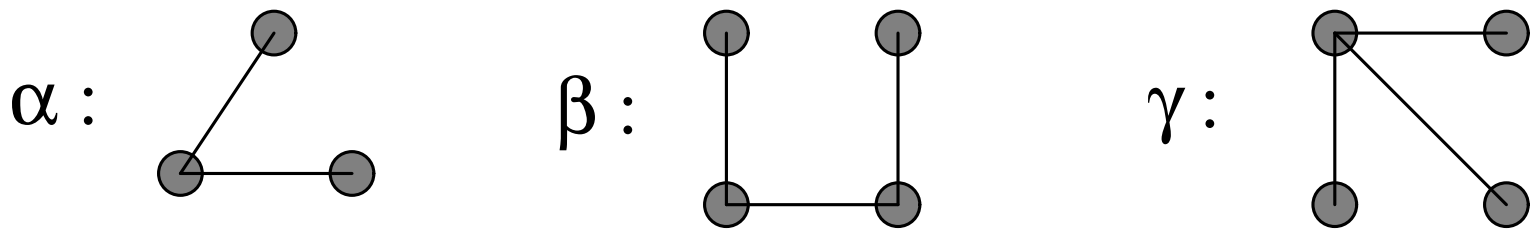

FIG. 12

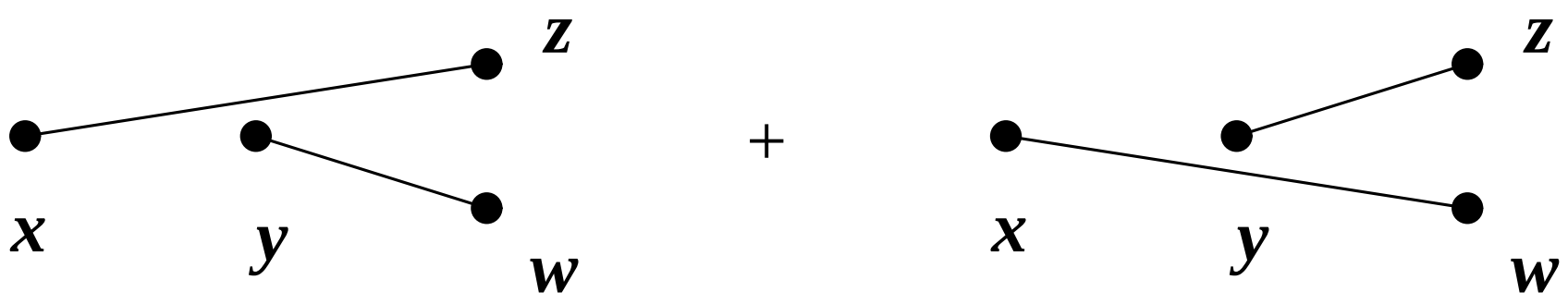

FIG. 13

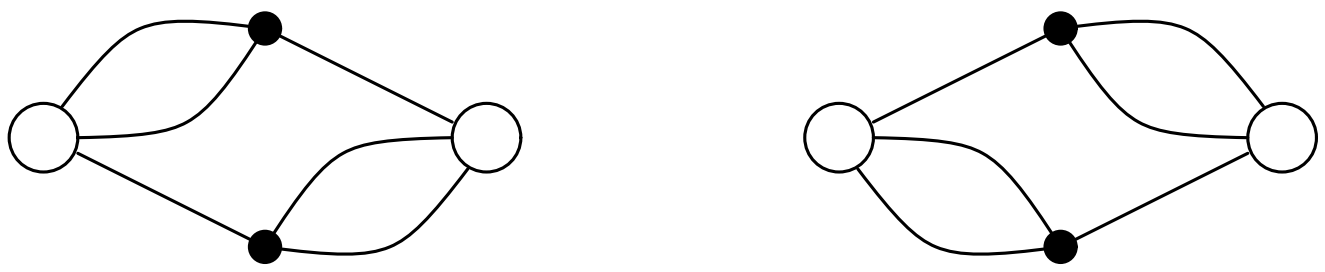